\documentclass[astrosymb,twocolumn]{aastex631}

\usepackage{gensymb}
\usepackage{lmodern}
\usepackage{amsmath}
\usepackage{natbib}
\usepackage{amssymb}
\usepackage{xfrac}
\usepackage{bm}
\usepackage{hyperref}
\usepackage[nameinlink]{cleveref}
\setcitestyle{citesep={;}}
\setcitestyle{aysep={}}

\creflabelformat{equation}{#2\textup{#1}#3}

\newcommand{\RNum}[1]{\uppercase\expandafter{{\scshape\romannumeral #1\relax}}}
\defcitealias{Rajpaul2015}{R15}
\defcitealias{Aigrain2012}{A12}

\shorttitle{Stellar Activity with GPs}
\shortauthors{Tran et al.}
\graphicspath{{./}{Figures/}}

\begin{document}

\title{Joint Modeling of Radial Velocities and Photometry with a Gaussian Process Framework}

\correspondingauthor{Quang H. Tran}
\author[0000-0001-6532-6755]{Quang H. Tran}
\email{quangtran@utexas.edu}
\affiliation{Department of Astronomy, The University of Texas at Austin, 2515 Speedway, Stop C1400, Austin, TX 78712, USA}
\affiliation{Center for Computational Astrophysics, Flatiron Institute, New York, NY}

\author[0000-0001-9907-7742]{Megan Bedell}
\affiliation{Center for Computational Astrophysics, Flatiron Institute, New York, NY}

\author[0000-0002-9328-5652]{Daniel Foreman-Mackey}
\affiliation{Center for Computational Astrophysics, Flatiron Institute, New York, NY}

\author[0000-0002-0296-3826]{Rodrigo Luger}
\affiliation{Center for Computational Astrophysics, Flatiron Institute, New York, NY}

\accepted{April 23, 2023}
\submitjournal{ApJ}

\begin{abstract}
    Developments in the stability of modern spectrographs have led to extremely precise instrumental radial velocity (RV) measurements. For most stars, the detection limit of planetary companions with these instruments is expected to be dominated by astrophysical noise sources such as starspots. Correlated signals caused by rotationally-modulated starspots can obscure or mimic the Doppler shifts induced by even the closest, most massive planets. This is especially true for young, magnetically active stars where stellar activity can cause fluctuation amplitudes of $\gtrsim$0.1 mag in brightness and $\gtrsim$100 m s$^{-1}$ in RV semi-amplitudes. Techniques that can mitigate these effects and increase our sensitivity to young planets are critical to improving our understanding of the evolution of planetary systems. Gaussian processes (GPs) have been successfully employed to model and constrain activity signals in individual cases. However, a principled approach of this technique, specifically for the joint modeling of photometry and RVs, has not yet been developed. In this work, we present a GP framework to simultaneously model stellar activity signals in photometry and RVs that can be used to investigate the relationship between both time series. Our method, inspired by the \textit{FF}$^\prime$ framework of \citep{Aigrain2012}, models spot-driven activity signals as the linear combinations of two independent latent GPs and their time derivatives. We also simulate time series affected by starspots by extending the \texttt{starry} software \citep{Luger2019} to incorporate time evolution of stellar features. Using these synthetic datasets, we show that our method can predict spot-driven RV variations with greater accuracy than other GP approaches.
\end{abstract}

\section{Introduction}
\label{sec:intro}

Instrumental advances in next-generation spectrographs are pushing the radial velocity (RV) method to new limits, facilitating the discovery of lower-mass planets around Sun-like stars. The RV method has been used to great success for the detection, confirmation, and characterization of exoplanets. Technological developments in the last decade, such as improved environmental control (e.g., temperature stabilization and vacuum seals), fiber feeds for better wavelength calibrations, increased spectral resolution, and broader wavelength coverage have led to higher fidelity spectra and lower instrumental errors \citep{Fischer2016}. Extremely precise instruments are now demonstrating measurement uncertainties at the tens of centimeters per second level in the optical and meter per second level in the near-infrared (NIR) \citep[e.g.,][]{Pepe2002, Mayor2003, Crane2010, Pepe2013a, Gibson2016, Jurgenson2016, Schwab2016, Donati2018, Seifahrt2018, Chakraborty2018, Quirrenbach2018, Metcalf2019, Pepe2021}.

As systematic instrumental effects are removed, planet detection limits are increasingly dominated by intrinsic stellar variability. This variability includes contributions from faculae, plages, and starspots \citep{Saar1997}; granulation \citep{DelMoro2004}; acoustic pressure (p-mode) pulsations \citep{Kjeldsen1995}; and long term magnetic activity cycles \citep{Santos2010}. Of these, starspots are particularly problematic. Starspots, or localized magnetic regions on the stellar surface, will suppress the local convective blueshift and appear cooler and redshifted compared to the surrounding photosphere. Dark spots will cause a relative drop in flux and perturb the spectral line profile as the star rotates \citep{Jahn1984, Donati1992, Schuessler1996}. Rotationally-modulated starspots will produce quasi-periodic variations in RV observations similar to those caused by the Doppler signals from planetary companions \citep{Aigrain2012}.

Intrinsic stellar variations are further intensified for young stars, making it even more difficult to discover and characterize planets. At younger ages, stellar rotation is faster, magnetic fields are stronger, and starspot coverage fractions are higher. These effects couple together, increasing stellar variability at young ages in time series observations. For example, at 5 Myr, light curve (LC) fluctuations and RV jitter can reach one mag and hundreds of meters per second, respectively, and completely envelope the signals of the closest, most massive planets \citep{Donati2016, Johns-Krull2016, Yu2017b, Rebull2020}. Even at older ages such as 50 Myr, activity signals can have photometric and RV amplitudes as high as 0.1 mag and $\sim$100 m s$^{-1}$, respectively \citep{Hillenbrand2015, Rebull2016, Rebull2020, Tran2021}.

However, planets around young stars provide critical information about planetary evolution including atmospheric escape, thermal evolution, and orbital migration \citep[e.g.,][]{David2016, Yu2017a, David2019, Plavchan2020, Rizzuto2020, Tran2021, Zakhozhay2022, Grandjean2023}. RV surveys have typically only targeted inactive, slowly rotating ($\lesssim$10 km s$^{-1}$), older stars ($\sim$1--10 Gyr) to avoid the effects of stellar activity. Thus, much of our knowledge about planetary evolution is constrained to after evolutionary processes such as radiative cooling, atmospheric loss, and dynamical interactions have already concluded. Investigating the differences in population demographics and architectures between young and old planetary systems will help reveal information on the mechanisms that drive planetary formation and migration pathways.

In order to detect and characterize planetary companions around younger, more active stars, we need to develop methods that can distinguish non-dynamical variations from Keplerian motion. This requires robust statistical techniques to simultaneously model the behavior of stellar activity and planet-induced signals in RV time series.

There have been a number of methods developed that attempt to do this. \citet{Dumusque2017} and more recently \citet{Zhao2022} presented data challenges to different precision RV teams, each with their own unique extraction process. They directly compare each framework's efficacy to recover planetary signals with known stellar activity contamination. Some approaches are described below, but for a more comprehensive list and description of each, see \citet{Dumusque2017} and \citet{Zhao2022}. Furthermore, a more general discussion of different approaches to mitigating stellar activity in RVs can be found in \citet{Hara2023}.

One technique has been to treat activity signals modulated by the rotation period as coherent sinusoids and simultaneously model them with the planet \citep[e.g.,][]{Boisse2011, Pepe2013b}. Another successful method is the floating chunk offset technique introduced by \citet{Hatzes2011}, which treats individual segments of the RV curve for activity offsets. The moving average approach does a similar analysis to fitting sinusoids in that it assesses correlations between groups of RV measurements with red noise filters \citep{Tuomi2013}. If spectral activity indicators are available, one can also fit sinusoids to those time series or linear trends to those indicators as a function of RVs to correct for any behavior \citep[e.g.,][]{Lovis2011, Dumusque2011, Meunier2013}. 

Of particular interest to this study are approaches that leverage photometric data to model stellar activity signals in RVs. As demonstrated by the Spot Oscillation And Planet \citep[SOAP;][]{Boisse2012, Dumusque2014} simulation framework, photospheric surface features like spots and plages are expected to influence both photometry and RVs, which motivates the joint fitting of both data sets in activity models. A notable example is LC inversion, which uses LCs to reconstruct the stellar surface map at the time of observation and predict RV variations driven by active regions. \citep{Lanza2010} adopted a maximum entropy map approach for the inversion process and synthesized the RV fluctuations of CoRoT-7. More recently, \citet{Roettenbacher2022} employed a similar algorithm and, in conjunction with interferometric data, reconstructed the surface map of $\epsilon$ Eri to produce an RV activity model. A more data-driven technique is the \textit{FF}$^\prime$ method \citep{Aigrain2012}, which defines a physically motivated relationship between fluctuations in photometry and RVs. That method has informed the modeling of time series observations for a number of advanced techniques.

One such advanced method is Gaussian process (GP) regression, which has emerged as a popular method to model the stochastic behavior of activity signatures in individual systems \citep[e.g.,][]{Haywood2014, Grunblatt2015, Dai2017}. In the particular case of activity signals in RV time series, GPs have been utilized in several different approaches. These techniques range in complexity, beginning with using a GP as a correlated red noise filter \citep[e.g.,][]{Baluev2013, Affer2016, Faria2020} to leveraging information from ancillary photometery to further constrain the GP trained on the RVs \citep[e.g.,][]{Haywood2014, Grunblatt2015, Dai2017}, to jointly modeling spectrosphic activity indicators and RVs with a multidimensional GP in order to constrain the activity signals \citep[e.g.,][]{Rajpaul2015, Jones2017, Gilbertson2020, Barragan2022}.

In this study, we present a GP framework that jointly models stellar activity signals in time series observations. Our framework utilizes two independent latent GPs and their time derivatives and primarily focuses on the joint modeling of RVs and photometry. The paper is structured as follows. We present a new method to generate synthetic starspot dominated time series observations using the \texttt{starry} package in \autoref{sec:stellar_activity}. We describe our joint GP framework in \autoref{sec:gp_framework}. In \autoref{sec:application}, we detail its application to simulated data and compare its performance to methods previously used in the literature. We discuss the implications of our framework and GPs in general in \autoref{sec:discuss}. Finally, we summarize our conclusions in \autoref{sec:summary}.

\section{Simulating Time Series Data}
\label{sec:stellar_activity}

We use the open-source software package \texttt{starry} \citep{Luger2019} to simulate light and RV curves of starspot-dominated stellar surface maps. The \texttt{starry} package treats the specific intensity map and the radial component of the velocity field of a star as sums of spherical harmonics. This decomposition into the spherical harmonic basis makes computing the total flux and RV integrated across the entire visible surface of the projected stellar disk possible to do analytically. In particular, \texttt{starry} has the capability to simulate stellar surface inhomogeneities, such as starspots, with arbitrary number, location, size, and contrast. The basic spot method in \texttt{starry} can inject spot features in the functional form of a circular top hat expanded in terms of spherical harmonics. This simple model can already facilitate the possibility of modeling the effects of starspots on photometric and RV curves.

We further develop the starspot model to better approximate spots of smaller sizes and simulate realistic starspot-dominated time series data. Spots in our framework are characterized by three descriptive parameters: contrast, radius, and time of maximum contrast, and two positional parameters: latitude and longitude. The spot contrast and radius, $c$ and $r$ respectively, describe the fractional change of the surface map intensity at the spot center and the radial size of the spot. Both $c$ and $r$ are largest at the time of maximum contrast, $t_0$. Spots are initialized as a 2D Gaussian on the surface of the map, centered at spot latitude, $\phi$, and longitude, $\lambda$. Spot contrast decays radially outward following the functional form: 
\begin{equation}\label{eq:spot_contrast_theta}
    c\,(\Delta \theta) = e^{-\frac{1}{2}(\frac{\Delta \theta}{r_0})^2},
\end{equation}
where $\Delta \theta$ is the angular distance away from spot center in degrees and $r_0$ is spot radius at time of maximum contrast.

We also implement spot emergence and decay to simulate the time evolution of spots. Spots are given a characteristic evolutionary rate with a functional form of:
\begin{equation}\label{eq:spot_evo}
    \eta (t) = e^{-\tau \left|t - t_0\right|},
\end{equation}
where $\tau$ is the evolutionary rate constant, in units of inverse days. The spot contrast and radius at some time $t$ are defined as the maximum spot contrast and radius multiplied by this evolutionary rate such that:
\begin{equation}\label{eq:spot_radius_time}
    r\,(t) = r_0 \eta(t).
\end{equation}
Specifically, spot contrast is set to evolve at twice the rate of the radius in order to ensure spot evolution is smooth at the time of maximum emergence and decay. Accounting for this effect and both radial (\autoref{eq:spot_contrast_theta}) and temporal dependence (\autoref{eq:spot_evo}), spot contrast is defined as:
\begin{equation}\label{eq:spot_evo_final}
    c\,(\Delta \theta, \, t) = e^{-\frac{1}{2}(\frac{\Delta \theta}{r_0})^2} \cdot c_0 e^{-2\tau \left|t - t_0\right|},
\end{equation}
where $c_0$ is the maximum spot contrast. \autoref{eq:spot_evo_final} can be used to construct a rotating \texttt{starry} surface map with evolving starspots. This is done by initializing a new \texttt{starry} map that has been rotated at each time step and the contrast for each starspot updated using \autoref{eq:spot_evo_final}. In addition to these spot parameters, \texttt{starry} requires several principal inputs in order to construct a stellar surface map. These components include $l_\mathrm{max}$, which encodes the maximum degree of the spherical harmonic expansion, and $i_*$ and $\Psi_*$, which correspond to the stellar inclination and obliquity, respectively, and define the orientation of the map relative to the viewer.\footnote{For a detailed tutorial on how to construct a stellar surface map and} add basic spot features onto a surface map, see the \texttt{starry} documentations at \href{https://starry.readthedocs.io/en/latest/notebooks/StarSpots/}{https://starry.readthedocs.io/en/latest/notebooks/StarSpots/}.

\begin{figure*}[!ht]
\centering
	\includegraphics[width=1\textwidth]{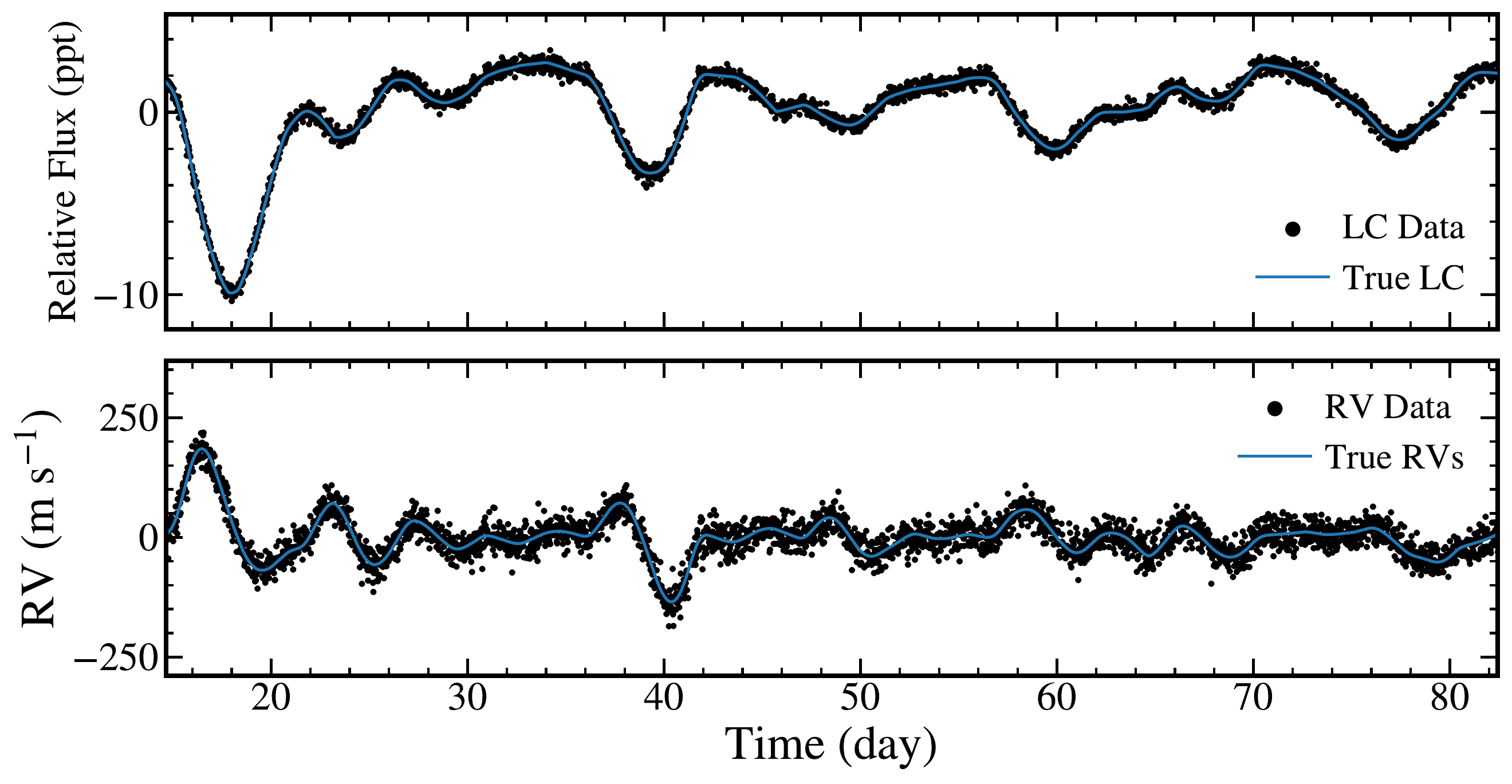}
    \caption{Synthetic photometric (top) and RV (bottom) observations generated using \texttt{starry} for a stellar surface map with $P_* = 11$ days ($v_\mathrm{eq} = 4.6$ km s$^{-1}$), $i_* = 86.5\degree$, $\Psi_* = 6.0\degree$, and $l_\mathrm{max} = 20\degree$. 50 spots each with a decay rate of $\tau = 0.05$ days$^{-1}$ and randomly distributed in longitude, latitude, contrast, radius, and time are placed on the surface of the map. Data mimic one quarter from the \textit{Kepler} space telescope with a 90-day baseline and 30 minute observational cadence. RV data are generated at the same cadence as the photometry, but this high sampling is not expected or required for the GP framework introduced in this work. Random scatter with an amplitude of 0.25 ppt for the LC and 20 m s$^{-1}$ for the RVs are injected to the true data (blue line), approximating uncertainties from both instrumental and astrophysical sources in order to simulate realistic, noisy observations of active stars for our tests (black dots).}
    \label{fig:starry_ex_data}
\end{figure*}

A representative set of time series data synthesized with 50 spots of random characteristics on a surface map with a spherical harmonic expansion degree of $l_\mathrm{max} = 20\degree$, a stellar inclination of $i_* = 86.5\degree$, and an obliquity of $\Psi_* = 6.0\degree$ is displayed in \autoref{fig:starry_ex_data}. Spot synthesis parameters are chosen in order to mimic realistic spots with sizes on the order of a few percent of the projected surface area and lifetimes close to one or several rotation periods. \texttt{starry} can generate both photometry and RVs at synthetically high cadence. Both time series in \autoref{fig:starry_ex_data} are displayed with observational cadence mimicking that of the \textit{Kepler} space telescope to demonstrate this capability and include injected measurement uncertainties of 0.25 ppt and 20 m s$^{-1}$ to the LC and the RVs, respectively, in order to simulate realistic time series noise of an active star.

\section{Joint Gaussian Process Framework}\label{sec:gp_framework}

\subsection{Gaussian Process Regression}\label{sec:gp_regression}

In this section we provide an introduction to Gaussian processes and GP regression in the context of modeling and analyzing time series data. We encourage readers interested in a more detailed discussion of GPs to review \citet{Rasmussen2006}.

A GP is a stochastic process where the joint prior distribution of the process outputs $\bm{y}$ are described by a multi-variate Gaussian distribution. The GP model consists of a deterministic mean function, $\mu(\bm{t}; \bm{\theta})$, and a covariance matrix or ``kernel'' function, $k(t_n, t_m; \bm{\alpha})$, which depend only on process inputs $\bm{t}$ and model parameters $\bm{\theta}$ and $\bm{\alpha}$. In our framework, the process outputs $\bm{y}$ are the time series observations at the corresponding input locations $\bm{t}$, defined as the times of observation. In the machine learning literature and in this work, the mean function is often assumed to be zero for simplicity, but it can be any continuous function with parameters $\bm{\theta}$ such as a sinusoid or Keplerian model. The covariance matrix describes the correlation between every pair of observation times, $t_n$ and $t_m$, based on some kernel function with parameters $\bm{\alpha}$. We discuss potential kernel function choices in \autoref{sec:kernels}.

Gaussian process regression leverages the property that observed data, which are treated as outputs or samples of a GP, are normally distributed in order to model stochastic processes with no known parametric expressions. As detailed above, the GP model is defined by the parameters of the mean function and the covariance matrix, \bm{$\theta$} and \bm{$\alpha$}. However, because our model outputs follow a Gaussian distribution, we are able to write down a log-likelihood function of the familiar form:
\begin{equation}\label{eq:gp_loglikelihood}
    \mathrm{log}\mathcal{L}(\theta, \alpha) = -\frac{1}{2}\bm{r}^\intercal \bm{K}^{-1} \bm{r} - \frac{1}{2}\mathrm{log}(\mathrm{det}\;\bm{K}) - \frac{N}{2}\;\mathrm{log}(2\pi),
\end{equation}
where $\bm{r} = \bm{y} - \mu(\bm{t}; \bm{\theta})$ is the residual vector of the data after subtracting the mean function, $\bm{K}$ is the covariance matrix whose elements are defined as $[\bm{K}]_{nm} = k(t_n, t_m; \bm{\alpha})$, and $N$ is the number of observations. The maximum \textit{a posteriori} (MAP) values of the mean function and covariance function parameters can be estimated by optimizing \autoref{eq:gp_loglikelihood}. Similarly, the joint posterior probability density of each parameter can be sampled using techniques such as Markov chain Monte Carlo (MCMC). Through these methods, GP regression can probe the underlying processes that influence observed correlations through physically informed mean and kernel functions.

\subsection{Model for Multiple Time Series Observations}
\label{sec:model}
\subsubsection{Physical Model}

GP regression can be especially useful in modeling correlated signals in different astronomical time series data. Since the variability seen in photometric and spectral measurements can originate from multiple sources, it is often difficult to distinguish between astrophysical versus dynamical effects with only one set of time series data. In previous studies, ancillary time series data have been used in GP regression to robustly treat stellar activity signals.

\citet[hereafter \citetalias{Aigrain2012}]{Aigrain2012} introduced a physically motivated framework that described the variations in photometry and RVs due to starspots. Their model established a simple relationship between the two time series and a formalism that predicts RV variations with only photometric information. \citetalias{Aigrain2012} first considered the effect caused by a single equatorial starspot. The relative drop in flux, $F(t)$, along the viewer line-of-sight due to the spot is equivalent to the projected area of the spot. As the spot rotates on the surface of the star, this projected area is dependent on the position of the spot, which varies as a function of the rotation period, or as cos$\:\phi(t)$ = cos$(2\pi t/P)$. Similarly, RV perturbations caused by activity, $\Delta \mathrm{RV}_\mathrm{rot}$, are attributed to the rotational modulation of the spot and therefore varies as a function of the rotational velocity, or as sin$\:\phi(t)$ = sin$(2\pi t/P)$. The model further accounts for effects from the suppression of convective blueshift, $\Delta \mathrm{RV}_\mathrm{c}$, by noting that it depends on spot position and thus similarly varies as a function of the rotation period. Both RV variation contributions additionally depend on the projected area of the spot. Thus, the total RV variation, defined as $\Delta \mathrm{RV} = \Delta \mathrm{RV}_\mathrm{rot} + \Delta \mathrm{RV}_\mathrm{c}$, can be re-expressed using a combination of the drop in flux due to the starspot, $F(t)$, and its time derivative $\dot{F}(t)$,
\begin{equation} \label{eq:A12_RV}
    \Delta \mathrm{RV}(t) = A_\mathrm{rot}F(t)\dot{F}(t) + A_\mathrm{c}F^2(t),
\end{equation}
where the coefficients $A_\mathrm{rot}$ and $A_\mathrm{c}$ are related to physical parameters but are treated as free variables in the model. This framework demonstrates that multiple time series data contain complementary information on the underlying stellar activity and can be used to construct an informed activity model. \citetalias{Aigrain2012} can provide a guide for a physically motivated framework to simultaneously fit both photometry and RV times series.

The relationship proposed by \citetalias{Aigrain2012}, denoted the \textit{FF}$^\prime$ method, is attractive for its low number of unknown parameters, but has several conditions that restrict its application. The primary limitation is the simplicity of the model. The \textit{FF}$^\prime$ relationship is derived from a model that assumes only a single point-like spot rotating across the stellar surface. This limits the \textit{FF}$^\prime$ method's ability to fully capture more complicated behaviors and structures, such as the effects of rapidly evolving spot emergence and decay, limb-darkening, overlapping families of spots, and more complex spot shapes. Furthermore, this technique requires values of both the flux and its time derivative. The flux time derivative cannot be directly measured and must be obtained by numerically differentiating the flux. In order to accurately constrain the behavior of the time derivative, the flux measurements must be high-precision and well-sampled. \autoref{eq:A12_RV} also only estimates RV variations that are contemporaneous with photometric observations, requiring the two datasets to be taken simultaneously. This can be somewhat difficult to achieve given realistic observational constraints. While requirements of potentially expensive observational constraints can be met with next-generation instruments, these conditions, coupled with its simple spot-model, restricts its usage to a limited number of situations.

Inspired by the \textit{FF}$^\prime$ method, \citet[hereafter \citetalias{Rajpaul2015}]{Rajpaul2015} introduced a framework that uses a single GP to jointly model RVs and other time series such as spectral activity indicators. \citetalias{Rajpaul2015} introduces an activity variable, defined as $F^2(t)$, which they describe as latent, or unobserved. Recall that in the \citetalias{Aigrain2012} formalism $F^2(t)$ is defined as the projected area of a single spot, or equivalently the relative drop in flux due to that spot, squared. This latent variable is described by a single GP, $G(t)$, such that $G(t) \equiv F^2(t)$. Therefore, in the \citetalias{Rajpaul2015} formalism, \autoref{eq:A12_RV} can be rewritten as
\begin{equation}\label{eq:R15_RV}
    \Delta \mathrm{RV}(t) = A_\mathrm{c}G(t) + A_\mathrm{rot}\dot{G}(t).
\end{equation}
\citetalias{Rajpaul2015} further defines similar relationships for variations in other spectral time series, such as the cross-correlation function (CCF) bisector inverse slope (BIS) and the chromospheric activity indicator log$\:R^\prime_\mathrm{HK}$. It is worth noting that \citetalias{Rajpaul2015} specified that their model is intended to be used with activity indices. Since these diagnostic measurements are extracted concurrently with the RVs from the same stellar spectra, they easily fulfill the requirement of contemporaneous observation. Furthermore, activity indicators such as log$\:R^\prime_\mathrm{HK}$ are closely related to the spot coverage fraction so their measurements are as similar to actual observations of the latent process, $G(t) = F^2(t)$, as possible.

However, a GP framework with only one latent process cannot jointly model photometry and RVs and account for every component considered in the \textit{FF}$^\prime$ method. Recall that in \autoref{eq:A12_RV}, the variable $F(t)$ refers to the \textit{drop in flux}, which is related to the actual measured quantity, the flux, or $\Psi(t)$, as
\begin{equation}\label{eq:A12_flux}
    F(t) = 1 - \frac{\Psi(t)}{\Psi_0},
\end{equation}
where $\Psi_0$ is a free parameter defined as the stellar flux in the absence of spots on the stellar surface. We can use \autoref{eq:A12_flux} to redefine \autoref{eq:A12_RV} in terms of what we actually measure, the flux, to obtain the general expression: 
\begin{equation}\label{eq:A12_RV_w_flux}
    \Delta\mathrm{RV}(t) = a + b\:\Psi(t) + c\:\Psi^2(t) + d\:\dot{\Psi}(t) + e\:\dot{\Psi}(t)\Psi(t),
\end{equation}
where free parameters such as $A_\mathbf{c}$, $\Psi_0$, and $A_\mathrm{rot}$ are absorbed into the coefficients $a$, $b$, $c$, $d$, and $e$. In this expression, there are four terms: $\Psi(t)$, $\Psi^2(t)$, $\dot{\Psi}(t)$, and $\dot{\Psi}(t)\Psi(t)$. A single GP and its time derivative can model, at most, the contribution of two variables, $\Psi(t)$ and $\dot{\Psi}(t)$ or $\Psi^2(t)$ and $\dot{\Psi}(t)\Psi(t)$, but not all four simultaneously. A minimum of two latent processes allows a GP framework to jointly model photometry and RVs in a compatible approach to the \textit{FF}$^\prime$ method.

Such a model is observationally complemented by the launch of the \textit{Transiting Exoplanet Survey Satellite} \citep[\textit{TESS};][]{Ricker2015}. Through \textit{TESS}, high-precision, high-cadence photometry of many nearby bright stars over a time baseline longer than the rotation period of active stars is now readily available. This new wealth of data provides a unique opportunity to further explore how stellar activity signals materialize in photometry. A principled GP method that can leverage this photometry is now needed.

\subsubsection{Gaussian Process Model}
\label{sec:gp_model}
Here, we describe our GP framework motivated by the \textit{FF}$^\prime$ method of \citetalias{Aigrain2012} and the approach introduced in \citetalias{Rajpaul2015}. In this model, which we call the two ``latent'' GP framework, there are two independent processes, $f(t)$ and $g(t)$, that together describe the stellar activity signals we can observe in photometry and RVs. As in \citetalias{Rajpaul2015}, the true values of the GPs in time are latent variables, which we do not directly infer or observe but rather marginalize over. These latent processes can be assumed to be GPs such that they describe stochastic data normally distributed about some mean function. We have two noisy observations, the flux, F,\footnote{This F is different from the \textit{F} used in \citetalias{Aigrain2012}, which refers to the \textit{relative drop in flux}, as defined in \autoref{eq:A12_flux}. Here, F refers to the measured stellar flux, which corresponds to $\Psi$ in \citetalias{Aigrain2012} (see their Equation 3).} with $N$ data points and the radial velocity variations, RV, with $M$ data points. These two time series are treated as linear combinations of the two latent processes and their time derivatives, $\dot{f}(t)$ and $\dot{g}(t)$, and are defined as:
\begin{equation}\label{eq:flux}
    \mathrm{F}(t_\mathrm{F}) = A\, f(t_\mathrm{F}) + B\, \dot{f}(t_\mathrm{F}) + C\, g(t_\mathrm{F}) + D\, \dot{g}(t_\mathrm{F}),
\end{equation}
\begin{equation}\label{eq:delta_rv}
\begin{split}
    \Delta\mathrm{RV}(t_\mathrm{RV}) = E\, &f(t_\mathrm{RV}) + H\, \dot{f}(t_\mathrm{RV}) \\ &+ I\, g(t_\mathrm{RV}) + J\, \dot{g}(t_\mathrm{RV}),
\end{split}
\end{equation}
where the coefficients $A$, $B$, $\cdots$, $I$, $J$ are free parameters that describe the amplitudes of the relationships between the two latent GPs, their time derivatives, and the two sets of observations. The leading coefficients $A$ and $E$ must always be positive to avoid degeneracies between the amplitudes. Here, the two time series can be observed at distinct times $t_\mathrm{F}$ and $t_\mathrm{RV}$, such that the two do not need to be measured in the same observation.

Note that by including two latent processes, this model is more generalized as compared to a direct translation of the \textit{FF}$^\prime$ model into GPs. The \textit{FF}$^\prime$ model is included as a special case if $f(t) \equiv F(t)$ and $g(t) \equiv F^2(t)$ ($F$ here refers to the drop in flux defined in \autoref{eq:A12_flux}) such that the RV variations, \autoref{eq:delta_rv}, become

\begin{equation}
\begin{split}
    \Delta \mathrm{RV}(t) = E\, &F(t) + H\, \dot{F}(t) \\ &+ I\, F^2(t) + J\, \dot{F}(t) F(t).
\end{split}
\end{equation}
This equation is equivalent to the expanded \textit{FF}$^\prime$ method expression for RV variations (\autoref{eq:A12_RV_w_flux}). We note that we keep the generalized model and do not enforce this special case, as squaring or taking the square root of a GP is computational difficult and often intractable. While we are inspired by the \textit{FF}$^\prime$ method, the increased flexibility afforded by these additional terms allows the general model to identify information shared between photometry and RVs that may be unaccounted for by more simple spot models. These possibilities include scenarios where the relationship is much more tenuous, such as when the two data sets are separated by a large gap in time.

Since any affine operator on GPs produces another GP \citep[i.e., the derivative of a GP and linear sum of GPs are GPs;][]{Rasmussen2006}, $\mathrm{F}$ and $\Delta \mathrm{RV}$, in \autoref{eq:flux} and \autoref{eq:delta_rv} respectively, are themselves both GPs. As such, we can write down the covariance function between any combination of the two. Following \citet{Osborn2010} and \citetalias{Rajpaul2015}, the covariance between two observations of $\mathrm{F}$ at any two time indices $t_\mathrm{F}$ and $t_\mathrm{F}'$ is 
\begin{equation}\label{eq:K11}
\begin{split}
    \bm{K_\mathrm{F,F}}(t_\mathrm{F}, t_\mathrm{F}') &= A^2\, k_f(t_\mathrm{F}, t_\mathrm{F}') + AB\, \frac{dk_f(t_\mathrm{F}, t_\mathrm{F}')}{dt_\mathrm{F}'} \\
    &+ BA\, \frac{dk_f(t_\mathrm{F}, t_\mathrm{F}')}{dt_\mathrm{F}} + B^2\, \frac{d^2k_f(t_\mathrm{F},t_\mathrm{F}')}{dt_\mathrm{F}dt_\mathrm{F}'} \\
    &+ C^2\, k_g(t_\mathrm{F}, t_\mathrm{F}') + CD\, \frac{dk_g(t_\mathrm{F}, t_\mathrm{F}')}{dt_\mathrm{F}'} \\
    &+ DC\, \frac{dk_g(t_\mathrm{F}, t_\mathrm{F}')}{dt_\mathrm{F}} + D^2\, \frac{d^2k_g(t_\mathrm{F},t_\mathrm{F}')}{dt_\mathrm{F}dt_\mathrm{F}'},
\end{split}
\end{equation}
where $k(t_\mathrm{F}, t_\mathrm{F}')$ is the covariance between two observations of a process at times $t_\mathrm{F}$ and $t_\mathrm{F}'$, $dk_f(t_\mathrm{F}, t_\mathrm{F}')/dt_\mathrm{F}'$ is the covariance between observations of a process at time $t_\mathrm{F}$ and its derivative at time $t_\mathrm{F}'$, $dk_f(t_\mathrm{F}, t_\mathrm{F}')/dt_\mathrm{F}$ is the covariance between observations of a process at time $t_\mathrm{F}'$ and its derivative at time $t_\mathrm{F}$, and $d^2k_g(t_\mathrm{F},t_\mathrm{F}')/dt_\mathrm{F}dt_\mathrm{F}'$ is the covariance between two observations of the derivative of a process at times $t_\mathrm{F}$ and $t_\mathrm{F}'$. These covariance relationships are further detailed in \autoref{sec:covariance_time_derivative}. The subscripts $f$ and $g$ refer to which of the two latent processes are considered. \autoref{sec:block_matrix_elements} provides the full expressions for the covariance functions between other combinations of the two observables (F and $\Delta$RV, $\Delta$RV and F, and $\Delta$RV and $\Delta$RV).

The final covariance between our multi-variate time series is a block matrix with the form

\begin{equation}\label{eq:block_matrix}
    \bm{\mathcal{K}} = \begin{bmatrix}
    \hyperref[eq:K11]{\bm{K_\mathrm{F,F}}} & \hyperref[eq:K12]{\bm{K_\mathrm{F,RV}}} \\
    \hyperref[eq:K21]{\bm{K_\mathrm{RV,F}}} & \hyperref[eq:K22]{\bm{K_\mathrm{RV,RV}}},
    \end{bmatrix},
\end{equation}
where each element in the block matrix is the corresponding covariance function between the two sets of observations. In particular, the off-diagonal elements, which encode the covariance between the photometry and the RVs, will typically always learn information regarding that relationship. This is true in almost all instances, excluding extreme scenarios where the two time series are fully independent or actually derive from a singular process.

\begin{figure}[!tp]
    \centering
    \includegraphics[width=1\linewidth]{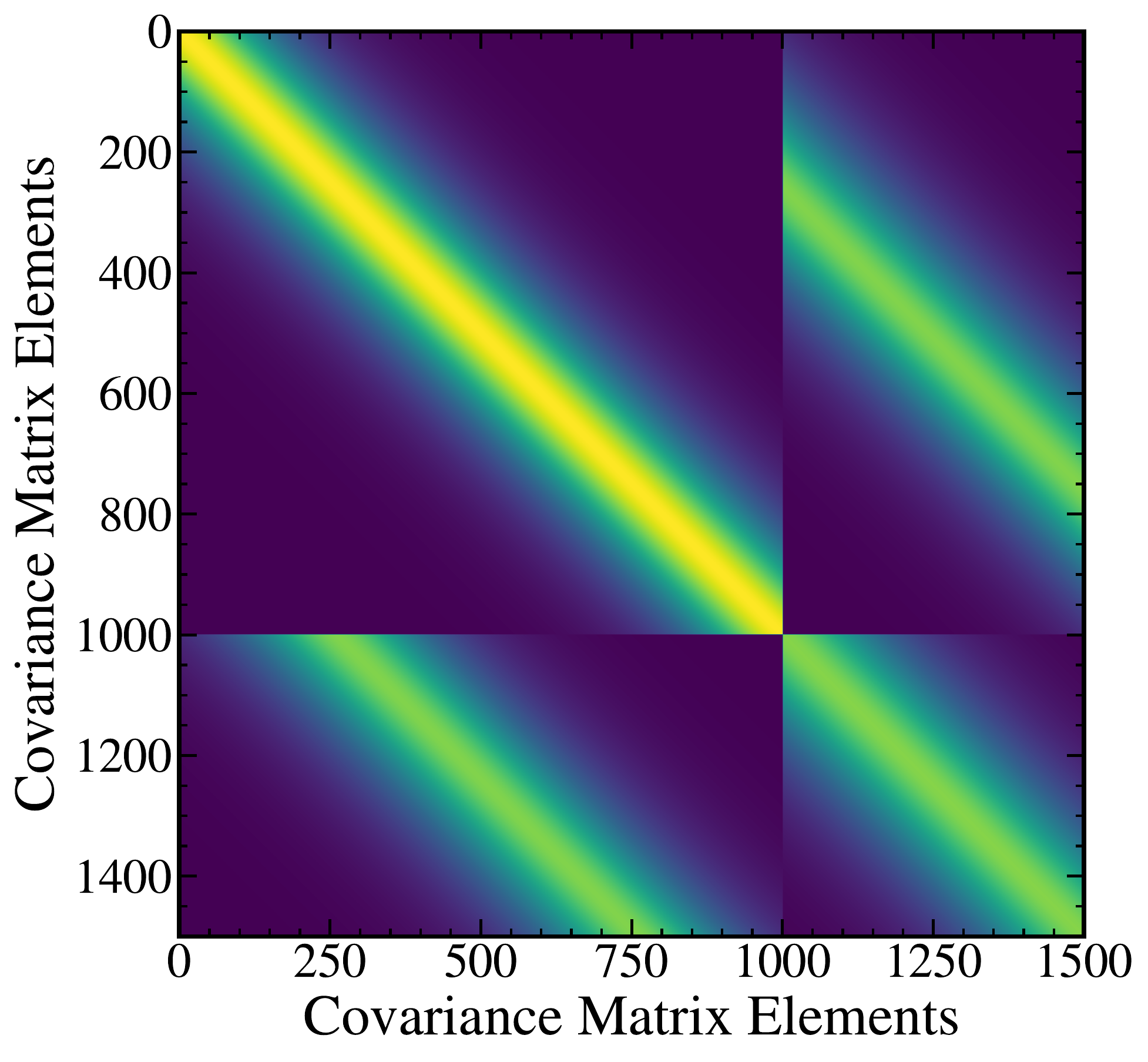}
    \caption{A representative covariance block matrix for our framework using a Mat\'ern--5/2 Kernel for a synthetic dataset of 1000 flux observations and 500 RV observations. The top left element is the covariance between flux observations (\autoref{eq:K11}), the off-diagonal elements are the covariances between flux and RV observations (\autoref{eq:K12} and \autoref{eq:K21}), and the bottom right diagonal element is the covariance of the RV observations (\autoref{eq:K22}).}
    \label{fig:matern52_cov_ex}
\end{figure}

\autoref{eq:block_matrix} is symmetric and positive semi-definite and is therefore a valid covariance matrix so long as each element of $\bm{\mathcal{K}}$ is a valid kernel function. If there are $N$ flux points and $M$ RV points, then $\bm{K_\mathrm{F,F}}$, $\bm{K_\mathrm{F,RV}}$, $\bm{K_\mathrm{RV,F}}$, and $\bm{K_\mathrm{RV,RV}}$ will be $N \times N$, $N \times M$, $M \times N$, and $M \times M$ matrices, respectively, and $\bm{\mathcal{K}}$ will be a $(N+M) \times (N+M)$ matrix. \autoref{fig:matern52_cov_ex} shows a representative covariance block matrix from our framework using a Mat\'ern--5/2 kernel function with a synthetic dataset of 1000 flux points and 500 RV points. 

We can now implement a GP regression (see \autoref{sec:gp_regression}) with our activity model. Given observations of F and $\Delta$RV at times $t_\mathrm{F}$ and $t_\mathrm{RV}$, we can create a total data vector $\bm{y}$ correspondingly observed at total time vector $\bm{t}$ by concatenating the two datasets (in the orientation of our framework, the flux dataset must be first). In the presence of planets, Keplerian and transit models can be incorporated as mean models for the RVs and flux. The covariance matrix can be constructed with any valid covariance kernel (see \autoref{sec:kernels}). With the residual vector $\bm{r}$ and covariance matrix $\bm{\mathcal{K}}$ in hand, a GP regression can be performed using \autoref{eq:gp_loglikelihood} to infer the model hyperparameters.

\subsection{Covariance Kernel Functions}
\label{sec:kernels}

The covariance kernel function is what determines the stochastic and periodic behavior of a GP model. Thus, selecting a kernel function that is both appropriate for the model and representative of the time series variations is imperative. Kernel selection entails balancing the flexibility required to encapsulate potential behavior of underlying processes while mitigating aggressive over-fitting of data. In particular, the covariances in our framework utilize the second time derivative of the kernel (see \autoref{eq:K11}), so the selected kernel \textit{must} be twice mean-square differentiable (i.e., both the kernel and its time derivative need to be smooth). There are a number of GP kernels that have previously been used to describe stellar variability in time series that meet this requirement. We summarize some potential kernel choices here, but for a more complete discussion on model selection, we refer the interested reader to \citet{Rasmussen2006}.

\subsubsection{Squared Exponential Kernel}
One of the most common kernel choice in GP inference is the squared exponential (SE), or Radial Basis Function (RBF), kernel, defined as

\begin{equation} \label{eq:sq_exp_kernel}
    k_\mathrm{SE}(t, t') = \sigma^2 \mathrm{exp}\left({-\frac{(t-t')^2}{2l^2}}\right),
\end{equation}
where $\sigma$ is the covariance amplitude and $l$ is the timescale of local correlations, or the growth and decay timescale of variations.\footnote{We define here the general form of the SE kernel but set $\sigma = 1$ in our framework as it is degenerate with the latent GP amplitudes.} The SE kernel is simple and infinitely differentiable, allowing it to produce smooth functions with no sharp discontinuities \citep{Duvenaud2014}. The SE kernel has been used to model correlated noise in both photometric and RV time series attributed to granulation \citep{Barclay2015}.

\subsubsection{Mat\'ern--5/2 Kernel}
The Mat\'ern family of kernel functions are another choice of kernels that are used to model less smooth behavior. The first of the Mat\'ern kernels that is twice differentiable is the Mat\'ern--5/2 (M$\sfrac{5}{2}$) kernel, that has the form
\begin{equation} \label{eq:matern52_kernel}
    k_{\mathrm{M}\sfrac{5}{2}} (t, t') = \left( 1 + x + \frac{x^2}{3} \right) e^{-x},
\end{equation}
where $x = \sqrt{5}\left|t-t'\right| / \rho$ and $\rho$ is the characteristic length scale. The M$\sfrac{5}{2}$ kernel (and its derivative) has been found to perform better than other GP kernels when tested on solar RV data \citep{Gilbertson2020}.

\subsubsection{Quasi-Periodic Kernel}
Of particular interest with regards to modeling stellar activity is the quasi-periodic (QP) kernel. The QP kernel has been previously employed based on the physical motivation that stellar activity signals caused by active regions on the surface of the star will modulate on the timescale of the stellar rotation period. Thus, as time series data often exhibit periodic or quasi-periodic behavior, this kernel has been widely used to constrain stellar activity signals in both photometry and RVs \citep{Aigrain2012, Haywood2014, Jones2017, Angus2018}. It is given by
\begin{equation} \label{eq:quasi_per_kernel}
    k_\mathrm{QP}(t,t') = A^2 \: \mathrm{exp}\left( -\frac{(t - t')^2}{2 l^2} - \frac{2 \: \mathrm{sin}^2\left(\frac{\pi (t - t')}{P}\right)}{\theta^2} \right),
\end{equation}
where $A$ is the covariance amplitude, $l$ is the local correlation timescale, $P$ is the recurrence timescale, or period, and $\theta$ is the smoothness of the periodic component, or importance of the periodic correlations.\footnote{The general form of the QP kernel uses the amplitude parameter $A$, but this is set to $A = 1$ in our framework due to its degeneracy with the latent GP amplitudes.} The quasi-periodic kernel is twice differentiable and, similar to the squared exponential kernel, produces functions that are very smooth.

\section{Application to Simulated Data}\label{sec:application}

In this section we apply the two latent GP framework to the fiducial dataset (displayed in \autoref{fig:starry_ex_data}) simulated using \texttt{starry} as described in \autoref{sec:stellar_activity}. The observational baseline of the photometric data considered covers $\sim$42 days of the total synthetic dataset, such that it is representative of data taken by \textit{Kepler} and \textit{TESS}. In order to simulate realistic observing conditions, the RV observational baseline is chosen to be half that of the LC ($\sim$21 days) with the time range inclusive of the photometric observations. When applying the GP regression, the LC is subsampled by a factor of five for an observational cadence of every 2.5 hours, totalling 400 data points. Similarly, 20 RV data points, randomly sampled, are used in the GP regression in order to mimic typical ground-based observational cadence (on average once-a-night).

\begin{figure*}[!th]
    \centering
    \includegraphics[width=1\linewidth]{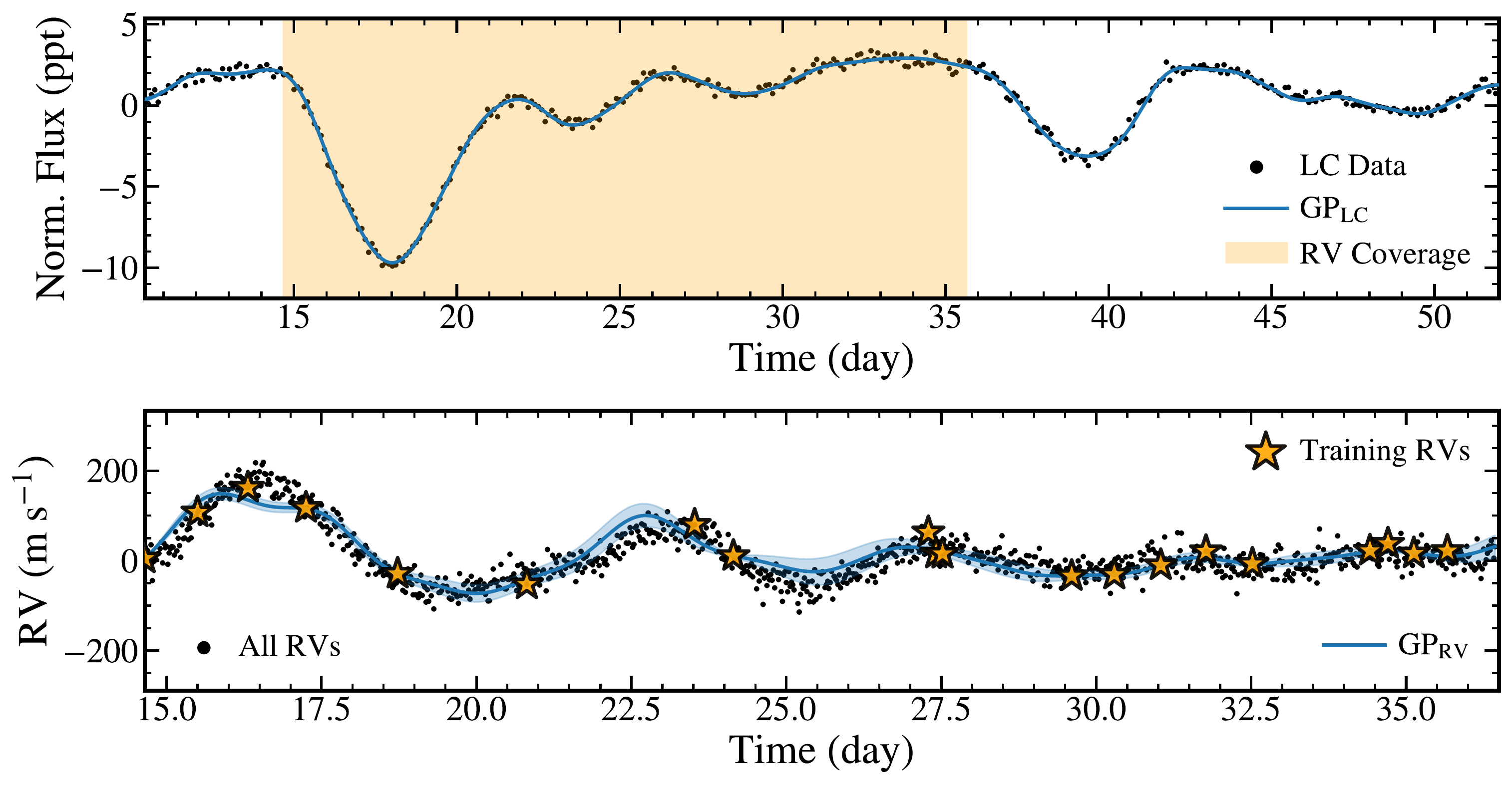}
    \caption{Application of the two independent latent GP framework using a quasi-periodic kernel on \texttt{starry} simulated photometric (top) and RV (bottom) observations. The synthetic data used here is a subset of the fiducial dataset shown in \autoref{fig:starry_ex_data} with different time baselines for each time series. All photometric data points with random noise added (top, black points) are used in the GP regression to simulate true photometric observations. The observational window overlapping between the photometry and the RVs is displayed as a shaded orange region in the top panel. The RV dataset is also shown (bottom, black points) but only 20 RV points (bottom, yellow stars) are used in the GP model regression. These 20 observations are randomly sampled to mimic realistic ground-based observation cadence. The best-fit GP model is shown in solid blue and the the 1$\sigma$ variance in shaded blue. The model performs well at predicting where the data lie even in gaps of coverage.}
    \label{fig:quasi_per_model_ex}
\end{figure*}

We apply a GP regression to the simulated dataset using each of the three covariance kernel functions described in \autoref{sec:kernels}. \autoref{fig:quasi_per_model_ex} displays the GP regression conditional predictions and 1$\sigma$ variances of our framework using the QP kernel. The same plots are displayed for the M$\sfrac{5}{2}$ and SE kernels in \autoref{sec:app_other_kernels}. In all three cases, cross-validation with the RV data points that are not used in training the GP model (black points) demonstrates the performance of the GP model at predicting the behavior of stellar activity. Through this comparison, we find that the three kernels perform similarly at reproducing the behavior of the fiducial dataset. We note that this is not an exhaustive list of all potential kernels. Any that meet our criteria can be used in the proposed latent GP framework.

\subsection{Comparison to Stellar Activity Models that Leverage Ancillary Photometry}
\label{sec:model_comparison}

There are several well known frameworks that can leverage ancillary photometry time series in order to model stellar activity signals in RVs. Each subsequent approach builds upon the last, further increasing the information content available to the GP regression and model flexibility. In this section, we briefly describe the methodologies of each method and compare their ability to predict the spot-driven variations in the simulated dataset against results of the two latent GP framework.

\subsubsection{\textit{FF}$^\prime$ Framework -- A Simple Spot Model}

As the two latent GP approach is inspired by the \textit{FF}$^\prime$ framework, it is therefore instructive to compare the results of this model when applied to the simulated fiducial dataset. We have already discussed the strengths and weaknesses of the \textit{FF}$^\prime$ model in \autoref{sec:model}. The efficacy of the \textit{FF}$^\prime$ technique in leveraging ancillary photometry to predict spot-driven RV variations is well-known. However, its simplicity restricts its application from more complex, and likely more realistic, spot distributions. In this application of \textit{FF}$^\prime$, we adopt the general expression defined in \autoref{eq:A12_RV_w_flux}.

\subsubsection{One GP Framework -- Red Noise Filter}
\label{sec:model_one_gp}

The simplest usage of GPs is directly training a single GP model on the RV time series, either with the GP alone or simultaneously with a Keplerian planet model \citep[e.g.,][]{Baluev2013, Affer2016, Faria2020}. This application of a GP treats the model as a correlated red noise filter. In the absence of prior knowledge about the nature of the stellar activity, this approach has several advantages. Due to the inherent flexibility of a GP, this approach will perform well at capturing the structure of the data. Furthermore, using a well motivated GP kernel leads to easily interpretable kernel hyperparameters that map to known physical processes. In the case of the QP kernel and photometry, many of these relationships are well-known \citep{Aigrain2012, Haywood2014, Angus2018}. However, a single GP trained only on RV data is prone to over-fitting the data. This is especially the case in sparsely sampled datasets where it is harder to test cross-validation and assess the predictive power of the GP model. Resolving this issue potentially requires obtaining many more expensive observations.

\subsubsection{Serial GP Framework -- Strong Priors}
\label{sec:model_serial}

An improvement to the simple red noise filter approach uses ancillary time series observations, such as photometry, to constrain the kernel hyperparameters. In this approach, a GP regression is first applied on the photometry and the inferred hyperparameter posteriors are adopted as strong priors for a GP model with the same kernel for the RVs (e.g., using the 3$\sigma$ highest density interval as the priors for each parameter). This method leverages the assumption that activity signatures are shared between time series data, motivated in part by the model from \citetalias{Aigrain2012}. As this approach serially trains the GP hyperparameters, first on the ancillary time series and then on the RVs, we call this the ``serial'' GP method. The serial GP framework provides additional safeguards against over-fitting the RV data by introducing an independent dataset containing additional activity information. It has been used successfully to model stellar activity using photometry as the ancillary data \citep[e.g.,][]{Haywood2014, Grunblatt2015, Dai2017}.

Despite its improvements, the serial GP approach can still fail to accurately capture the activity signals in RVs. As starspot-driven signals typically evolve on the timescales of the stellar rotation period, the ancillary time series observations should ideally be contemporaneous with the RVs in order to fully leverage additional information about stellar activity. This requirement can be prohibitive given observational constraints. Moreover, there is evidence that activity parameters are not explicitly shared between the photometry and RV signals \citep{Kosiarek2020, Nicholson2022}.

\subsubsection{One Latent GP Framework -- Multidimensional GPs}
\label{sec:model_one_latent}

GPs have also been used in multidimensional cases \citepalias[e.g.,][]{Rajpaul2015}. As previously described in \autoref{sec:model}, in this framework, stellar activity signals in RVs and multiple spectral activity indicators are modeled with linear combinations of a single GP and its time derivative. We call this approach the one latent GP method since the GP is not directly trained on each time series as in previous frameworks. Instead, the kernel hyperparameters are implicitly constrained by fitting each separate dataset to its respective relationship to the GP. For example, \citetalias{Rajpaul2015} defines the RV variation due to activity as a sum of the GP and its time derivative while the chromospheric activity index log$\:R^\prime_\mathrm{HK}$ is approximated using only the GP. Several variations of this framework have been proposed, including the use of multiple time derivatives of the latent GP and generic activity indicators instead of specific traditional ones \citep{Jones2017, Gilbertson2020, Barragan2022}.

The one latent GP approach improves on multiple aspects of traditional GP applications. In this framework, the time derivative(s) of the GP are used in the modeling. Adding a GP time derivative term addresses many of the flexibility issues with the serial GP case as it can capture phase shifts, harmonic complexity differences, and general asynchronous behavior between different time series \citep{Barragan2022}. The latent nature also allows for the simultaneous modeling of multiple time series observations, so long as a relationship between each time series and the underlying GP is defined. The addition of activity indicators as ancillary sources of information ensures that time series observations are measured at the same time and by the same instrument as the RVs.

Despite the many advantages of the one latent GP framework, there are still areas where improvements can be made, particularly for the joint modeling of photometry and RVs. While the model flexibility is increased by modeling the GP as a latent variable and including the time derivative as an additional term, the one latent GP technique can still be too restrictive. The model flexibility can be improved upon by adopting additional terms such as higher derivative terms \citep[e.g.,][]{Jones2017, Gilbertson2020}. This is especially true in scenarios that we discussed previously in \autoref{sec:model}. In instances where only photometric measurements are available, the one latent GP model will likely be too constrained by the LC information due to the typically multiple orders of magnitude more data points. This will cause the underlying GP model to be highly constrained by the LC features. Since these features are not necessarily shared with the RVs, this can actually decrease the accuracy with which the GP framework can predict activity signals in RVs.

\subsubsection{Model Performances \label{sec:model_performances}}

We evaluate the performance of our two latent GP framework and compare the results to the four other approaches described above. As the QP kernel is frequently chosen to model stellar activity, we adopt it for all subsequent GP comparisons. This decision reduces the effect of kernel choice on differences in performance between the GP models. However, tests were also performed with other kernels and the results are consistent across different kernels. All methods are applied to the same subsampled fiducial dataset described in \autoref{sec:application}, using all available information for that model, i.e., the \textit{FF}$^\prime$, serial, one latent, and two latent GP frameworks all use the photometry and RVs while the one GP framework use only the RVs. In the instance of the \textit{FF}$^\prime$ model, we follow the prescription from \citetalias{Aigrain2012} and interpolate the subsampled LC in order to calculate the flux and flux derivative values at the exact time stamps of the RV observations. Techniques are evaluated through their root-mean-square (RMS) between RVs predicted by the model and the fiducial RVs, not including the data used to train the model regression. This test is equivalent to cross-validation and is able to assess the accuracy of each model's predictive power. We note again that the fiducial dataset has an injected random scatter of 20 m s$^{-1}$ to represent instrumental noise contributions. In theory, a perfect prediction of the RV variations will still only reach this noise floor.

\begin{figure*}[!t]
    \centering
    \includegraphics[width=1\linewidth]{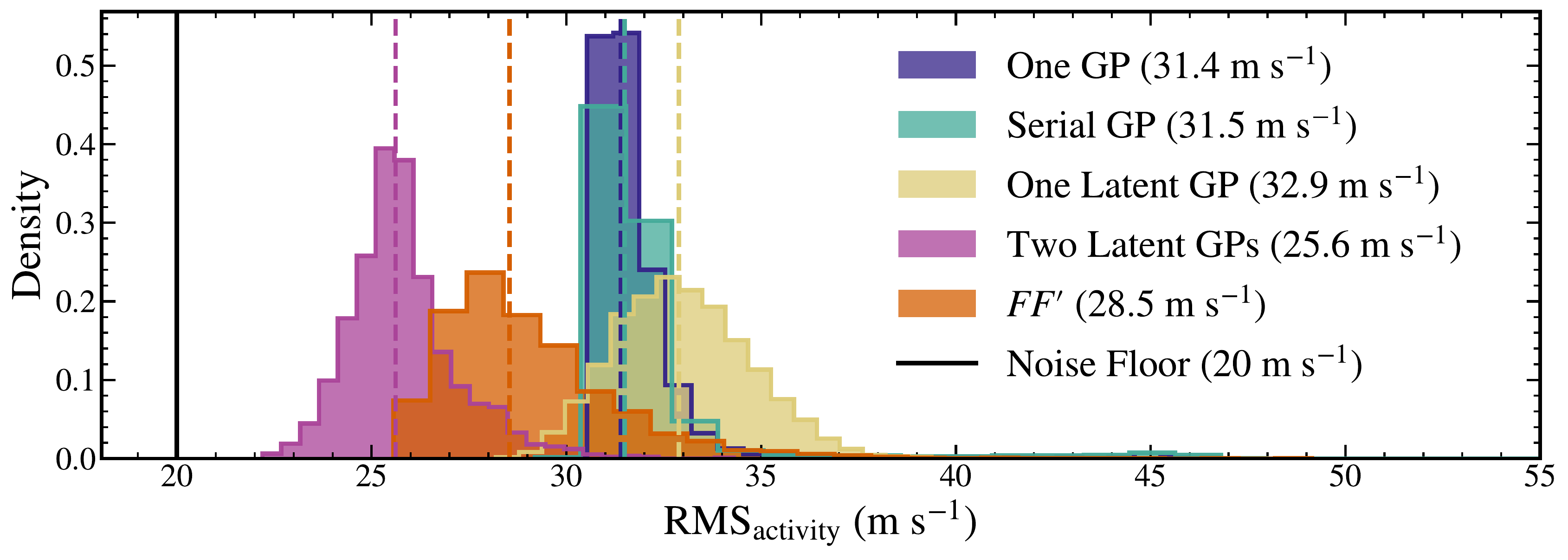}
    \caption{RMS distributions calculated from predicted models using parameter posteriors for each of the GP approaches. The \textit{FF}$^\prime$, one, serial, one latent, and two latent GP frameworks are shown in orange, blue, green, yellow, and purple, respectively. A solid black line denotes the injected 20 m s$^{-1}$ noise floor. The median RMS values for the one and serial GP models are similar (RMS $\approx$31 m s$^{-1}$), suggesting that additional photometric information does not necessarily increase the predictive power of a GP. The one latent GP framework performs, on average, the worst of the five models (RMS = 32.9 m s$^{-1}$). Under these ideal synchronous observational conditions, the \textit{FF}$^\prime$ model outperforms the one, serial, and one latent GP models (RMS 28.5 m s$^{-1}$). Our two latent GP framework performs the best, with a median RMS value of 25.6 m s$^{-1}$. Using the same amount of information as the \textit{FF}$^\prime$, serial, and one latent GP frameworks, our two latent GP model decreases the difference to the synthetic noise floor by approximately a factor of 2 as compared to other models.}
    \label{fig:rms_comp}
\end{figure*}

We sample model hyperparameters for each of the five frameworks using the \texttt{emcee} open software package \citep{Foreman-Mackey2013, Foreman-Mackey2019b}. We use 100 walkers and sample 10000 times for every model for a total of $10^6$ posterior samples. We adopt uninformative (uniform) priors for the free parameters of every model, except the serial model. This includes an additional white-noise jitter term, $\sigma_\mathrm{jit}$, that is added to the likelihoods. The priors for all the \textit{FF}$^\prime$ model coefficients are $\mathcal{U}[-250, 250]$ and the prior for the white-noise term is $\sigma_\mathrm{jit} = \mathcal{U}[\mathrm{log}(0.01), \mathrm{log}(100.0)]$. The priors on the QP kernel parameters are $l = \mathcal{U}[0.1, 30]$, $\theta = \mathcal{U}[0.1, 30]$, $P = \mathcal{U}[1.1, 30]$, and $\sigma_\mathrm{jit} = \mathcal{U}[\mathrm{log}(10^{-8}), \mathrm{log}(0.1)]$. The priors for the GP model parameters are all set to $\mathcal{U}[-250, 250]$ or $\mathcal{U}[0.01, 250]$, if it is the first coefficient. The serial GP uniform priors are bounded by the 3$\sigma$ highest density interval range of the parameter posteriors sampled from the LC. An RV curve prediction is generated using the last 5000 posterior samples and corresponding RMS values are calculated. Through this procedure, we create RMS posterior distributions for each framework. We can compare these posterior RMS distributions to see how each model performs at predicting RV variations. \autoref{fig:rms_comp} displays the five RMS distributions. The \textit{FF}$^\prime$, one, serial, one latent, and two latent GP approaches are shown in orange, blue, green, yellow, and purple, respectively. The median values of the RMS distributions are listed and displayed as dashed vertical lines.

For this test, we find that our framework improves fit residuals with respect to the injected noise floor by approximately a factor of two as compared to other approaches. An initial expectation may be that as we introduce ancillary information (photometry), stricter prior constraints (pre-conditioning on photometry), or additional model complexity (additional model parameters or GPs), the accuracy of the model's predictive power should increase. Instead, we do not necessarily observe this trend. For this fiducial dataset, the one and serial GP models have similar median RMS values (31.4 and 31.5 m s$^{-1}$, respectively) and overlapping RMS distributions. We also observe that the one latent GP framework performs the worst of the four, with a median RMS value of 32.9 m s$^{-1}$. For this ideal scenario where the ancillary photometry is measured with high cadence and within the same observational time window, the \textit{FF}$^\prime$ model performs better than the other three GP methods, with a median RMS value of 28.5 m s$^{-1}$. The two latent GP model does the best at capturing the behavior of starspot-driven activity, with a median RMS of 25.6 m s$^{-1}$. This is approximately a factor of two decrease in RMS relative to the synthetic noise floor of the RV dataset.

These outcomes can be understood in context of the limitations of each approach. The serial GP performing similarly to the one GP suggests that the former did not increase predictive power by leveraging the additional photometric information. This aligns with previous results indicating a disparity between GP hyperparameters for RVs and photometry. Furthermore, the serial model posteriors become fairly nonrestrictive. Despite expectations that the photometric posteriors would impose much more stringent constraints on the GP parameters, for at least two of the parameters ($l$ and $P$), the pre-conditioned priors did not appreciably change and were essentially the full range of the broad uniform priors (see \autoref{sec:model_one_gp}). The third parameter, $\theta$, did tighten to approximately one-fourth of the original parameter boundaries, but this is still moderately broad enough to permit the GP model ample freedom. In this case, the serial GP model becomes similar to the one GP approach and does not increase the additional predictive power to the model by much. Thus, that the performance of both models are similar is not unexpected.

The one latent GP approach performing the worst may seem contradictory to expectations but ultimately coincides with the caveats of the model listed in \autoref{sec:model_one_latent}. In the fiducial training set, the photometry data has a factor of 20 times more data points. Due to this disparity, the one latent GP model is primarily restricted to characteristic features in the LC. Here, the inconsistencies in hyperparameters between the photometry and RVs propagate forward in the modeling. The one latent GP regression over-weights the photometric data and predicts a model that captures the LC behavior well but RV variations poorly. For this fiducial dataset, the use of only photometry as the ancillary dataset leads to worse performance. We note that we find that this behavior will not happen in every circumstance.

Finally, the excellent performance of the \textit{FF}$^\prime$ model over the other three GP models is not unexpected given the observational design of the comparison. The LC and RVs in this instance perfectly overlap. Furthermore, despite the subsampling of the LC, the photometric data points are still observed at a high enough cadence that interpolation is possible. This allows for a robust numerical estimate of the flux derivative. These are the exact conditions for which the \textit{FF}$^\prime$ method is designed. In different observational instances, the \textit{FF}$^\prime$ technique may be less realistically achievable and therefore may not perform as well. However, even in this ideal scenario, the simplistic spot model of \textit{FF}$^\prime$ is likely limiting its ability to fully capture the RV variations caused by the complex and rapidly evolving starspot distributions on the surface map.

\begin{figure*}[!t]
    \centering
    \includegraphics[width=1\linewidth]{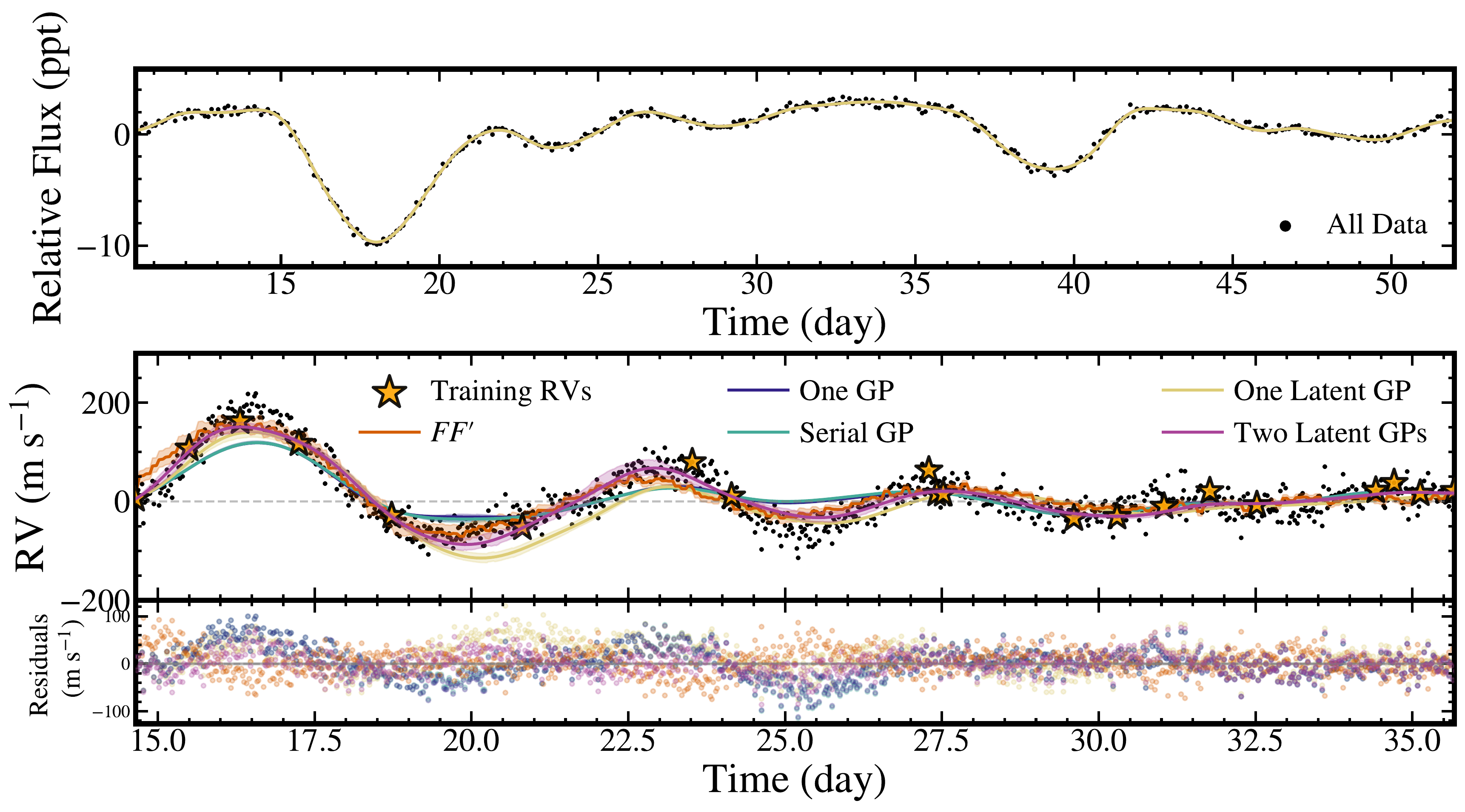}
    \caption{Mean and 1$\sigma$ highest density intervals of the predicted posterior models for each of the five approaches. The photometric and RV times series used (black points) are the same as in \autoref{fig:quasi_per_model_ex}. The colors for each framework are the same as in \autoref{fig:rms_comp}. All three model fits on the photometric data (serial, one latent, and two latent) are shown but cannot be seen as they are directly on top of each other. The one GP and serial GP models are very nearly on top of each other. The one latent GP prediction appears to have a slight phase shift from the RV points. The \textit{FF}$^\prime$ prediction has more high frequency variations, which are likely due to numerical approximations and interpolation schemes adopted to calculate the flux gradient. The two latent GP approach performs the best at capturing the quasi-periodic structure in the RV curve.}
    \label{fig:model_predict}
\end{figure*}

These trends are evident when examining the predictive models generated from posteriors of each framework. \autoref{fig:model_predict} shows the RV dataset, the average predictive model (solid lines) and the 1$\sigma$ standard deviations (shaded regions). Predictive models are generated using the average parameters of the model posteriors. Standard deviation contours are estimated by taking the 1$\sigma$ highest density interval of the average model predictions at each individual timestamp. The colors for each framework are the same as in \autoref{fig:rms_comp}. The black points are the full RV data and the yellow stars represent the RV training set.

The two latent GP framework does the best of tracking the quasi-periodic features of the RV curve. The one and serial GP approaches cannot be distinguished because they overlap one another. The one latent GP model fails to capture the RV points by a slight offset, possibly reflecting the phase shift of the photometry from the RVs. The standard deviation of the one latent GP prediction is plotted, but cannot be seen. In this case, the one latent model is extremely confident that it is able to predict the RV data. However, this confidence is attributed to the high number of photometric points and does not reflect true predictive power of the RV variations. The \textit{FF}$^\prime$ prediction follows the general variational trend well, but does not capture the full extent of the amplitude changes. There are also high-frequency fluctuations that are seen in the mean prediction that are likely due to the interpolation of the subsampled LC and approximations made in order to calculate the derivative of the flux.

\subsection{Non-simultaneous Photometric Coverage \label{sec:non_simultaneous}}

Here, we conduct a similar test as above to assess model performance on datasets with RVs both synchronous and asynchronous with the photometric observational window. This comparison is another, more realistic test of model predictive power.

\begin{figure*}[!t]
    \centering
    \includegraphics[width=1\linewidth]{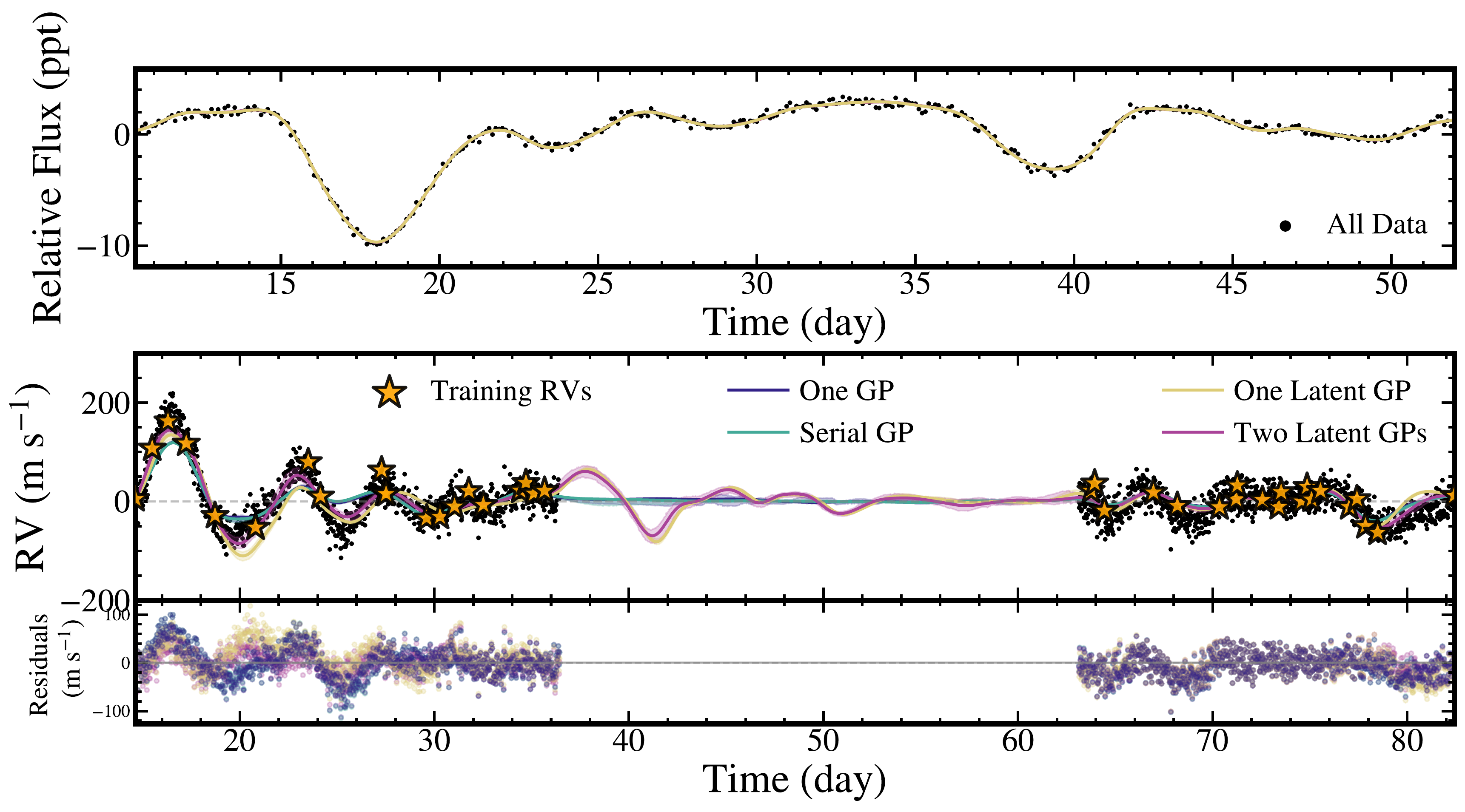}
    \caption{Mean and 1$\sigma$ highest density intervals of the predicted posterior models for each of the four GP approaches on the extended fiducial dataset including asynchronous RVs. Colors and symbols are the same as in \autoref{fig:model_predict}. Predictions for each model behave similarly in the synchronous RV and photometry time window. No data is plotted in the gap between RV observations but model predictions are presented. Notably, both the one latent and two latent predicted structure are similar to the actual behavior of the data (see \autoref{fig:starry_ex_data}).}
    \label{fig:model_predict_offset}
\end{figure*}

In the following comparisons, we include an additional asynchronous set of RV data points. These additional RVs have the same time baseline and sampling cadence as the contemporaneous RVs used above, but are separated by an ``observational gap'' from the original fiducial RVs and the photometric time series by one and two rotation periods (11 and 22 days), respectively (see \autoref{fig:model_predict_offset}). For this non-simultaneity experiment, sufficient separation between observation windows is needed to avoid potential effects from persistent stellar signals due to phase shifts between variations in photometry and RVs (see \autoref{sec:model_serial}). In this case, a separation of 11 days allows us to limit information between the RV observations while a gap of 20 days minimizes the persistent signals shared between the photometry and the new asynchronous RVs.

\begin{figure*}[!t]
    \centering
    \includegraphics[width=1\linewidth]{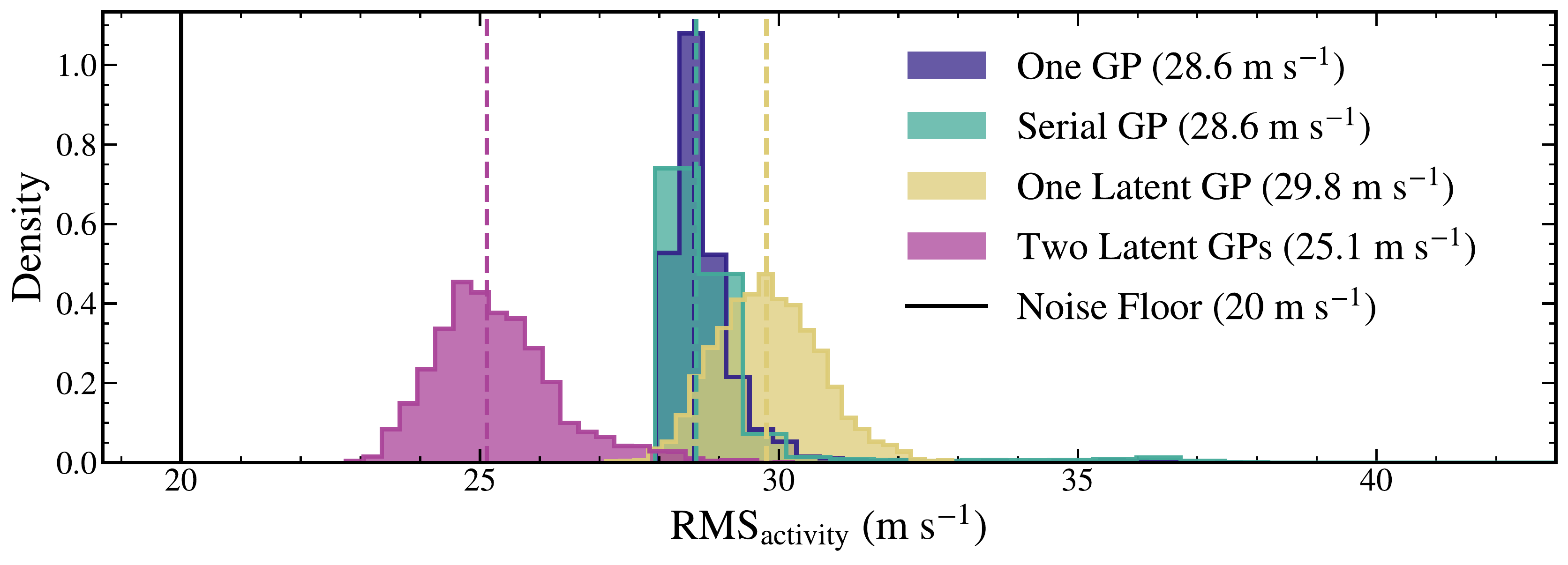}
    \caption{RMS distributions for predictions of each GP approach on the extended fiducial dataset including asynchronous RVs. The colors correspond to the same models as in \autoref{fig:rms_comp}. Model performances are consistent with previous synchronous tests. This suggests that the different approaches utilize the same activity information content from photometry in both instances of simultaneous and non-simultaneous RV coverage. That the two latent framework performs the best in this test supports its ability to best leverage information from asynchronous ancillary photometry.}
    \label{fig:rms_comp_offset}
\end{figure*}

We apply all four GP frameworks to this extended set of observations, which includes both synchronous and asynchronous RVs. \autoref{fig:rms_comp_offset} displays the RMS distributions for each approach on this new dataset and \autoref{fig:model_predict_offset} presents the corresponding mean and standard deviation model prediction. In order to check whether these results are affected by sampling between the synchronous and asynchronous RVs, we also apply this test for the case where photometric coverage is simultaneous with the second extended RV section instead of the first segment. For this alternative arrangement, we find that the results do not appreciably change.

The resulting model performances are consistent with the previous synchronous test. The one and serial GP approaches overlap with each other (RMS $ = 28.6$ m s$^{-1}$). Examining the photometry posteriors in the serial GP case again reveals that they are again within 10\% of the broad uniform priors imposed on the one GP model. As before, this reduces the serial GP approach to the one GP framework. Our two latent model and the one latent GP model continue to perform the best and worst, respectively. This behavior may be driven primarily by the asynchronous RV observational window, where there is a large decrease in variation amplitude. Thus, each model can more easily predict the low variations of the RVs despite there being no overlapping photometry to provide additional information on activity signatures. This is supported by the overall increase in performance despite the addition of new data, suggesting that all models are capable of predicting the behavior of the second asynchronous RV observational window.

The behavior exhibited in the gap between RV coverage (\autoref{fig:model_predict_offset}) provides additional insight on how each model leverages the ancillary photometry. No data is plotted in this separation while the best-fit model fits are still shown. Here, we see the one and serial GP approaches lose all predictive power where there are no data, especially in this instance where the training RV data points are relatively sparsely sampled and the gap in coverage is large. Alternatively, the latent GP models anticipate additional RV fluctuations, in particular at the beginning of the RV gap and while the photometric data continues to be observed. The latent GP frameworks are similar and model the actual data structure well (see \autoref{fig:starry_ex_data}). The differences between the two predictions are consistent with the synchronous period, where the one latent GP model likely demonstrates the known phase offset between RVs and photometry. As soon as the photometry coverage ends, the variance of the one latent GP model increases, further indicating the large dependence on the model on the high cadence photometry.

\subsection{Simulated Planet Injection and Recovery}

In this section, we provide a simple demonstration on how to jointly fit both a Keplerian and an activity model with the two latent GP framework. We caution that this preliminary example is not a definitive comparison between the efficacy of the two latent GP model in predicting stellar activity signals or in recovering planetary signals relative to other GP methods. Instead, the former comparison is explored in limited test cases as described in \autoref{sec:model_comparison}, while the latter comparison requires extensive tests using both simulated and real datasets applied to all methodologies discussed herein. Those comparisons are a natural extension of this introductory work. The following example is a demonstration of the efficacy of the two latent GP model for particular systems of interest such as close-in giant planets around young, active stars (see \autoref{sec:young_stars} for more details).

Here, we inject a synthetic Keplerian signal into the fiducial RVs and perform a simultaneous fit of both the two latent GP stellar activity model and a planetary model using the training dataset (subsampled photometry and sparse RVs). The simulated planet is defined with 6 parameters: the orbital period ($P_\mathrm{orb}$), the planetary mass ($M_p$), the time of periastron passage ($T_0$), the orbital inclination ($i_*$), and the re-parameterized versions of eccentricity and argument of periastron ($\sqrt{e}\;\mathrm{sin}\;\omega$ and $\sqrt{e}\;\mathrm{cos}\;\omega$). Given the fluctuation amplitudes in the fiducial RVs and expected stellar jitter of young, active stars, we chose to inject a hot Jupiter, which should produce an RV semi-amplitude similar to that of the stellar activity-driven signals.

The posteriors for the five Keplerian model parameters (including RV semi-amplitude and excluding stellar inclination) and 14 activity model parameters (8 model coefficients and 6 kernel hyperparameters) are sampled using the \texttt{emcee} open software package \citep{Foreman-Mackey2013, Foreman-Mackey2019b}. We impose non-informative uniform priors on every parameter, use 100 walkers and 50000 steps, for a total of $5 \times 10^6$ posterior samples, and remove the first 50\% of the posteriors as burn-in. \autoref{tab:planet_params} reports the imposed priors, true injected values, recovered median, and $\pm1\sigma$ highest-density interval (HDI) for each of the fitted planetary parameters.

We find that the joint Keplerian and stellar activity model recovers the true orbital parameters of the injected planet within one-sigma uncertainties. The residual RMS of the model constructed with the median values reported in \autoref{tab:planet_params} is 25.4 m s$^{-1}$. This residual scatter is comparable to the RMS predictions of the two latent GP model regressed on the stellar activity alone (\autoref{fig:rms_comp}). Furthermore, the injected signal is recovered at a reasonable significance $(K_p / \sigma \gtrsim 3.5)$ despite the low number of RV observations ($N_\mathrm{RV} = 20$) and the RV semi-amplitude of the simulated planet (70.5 m s$^{-1}$) being near or below the activity level.

\begin{figure*}[!t]
    \centering
    \includegraphics[width=1\linewidth]{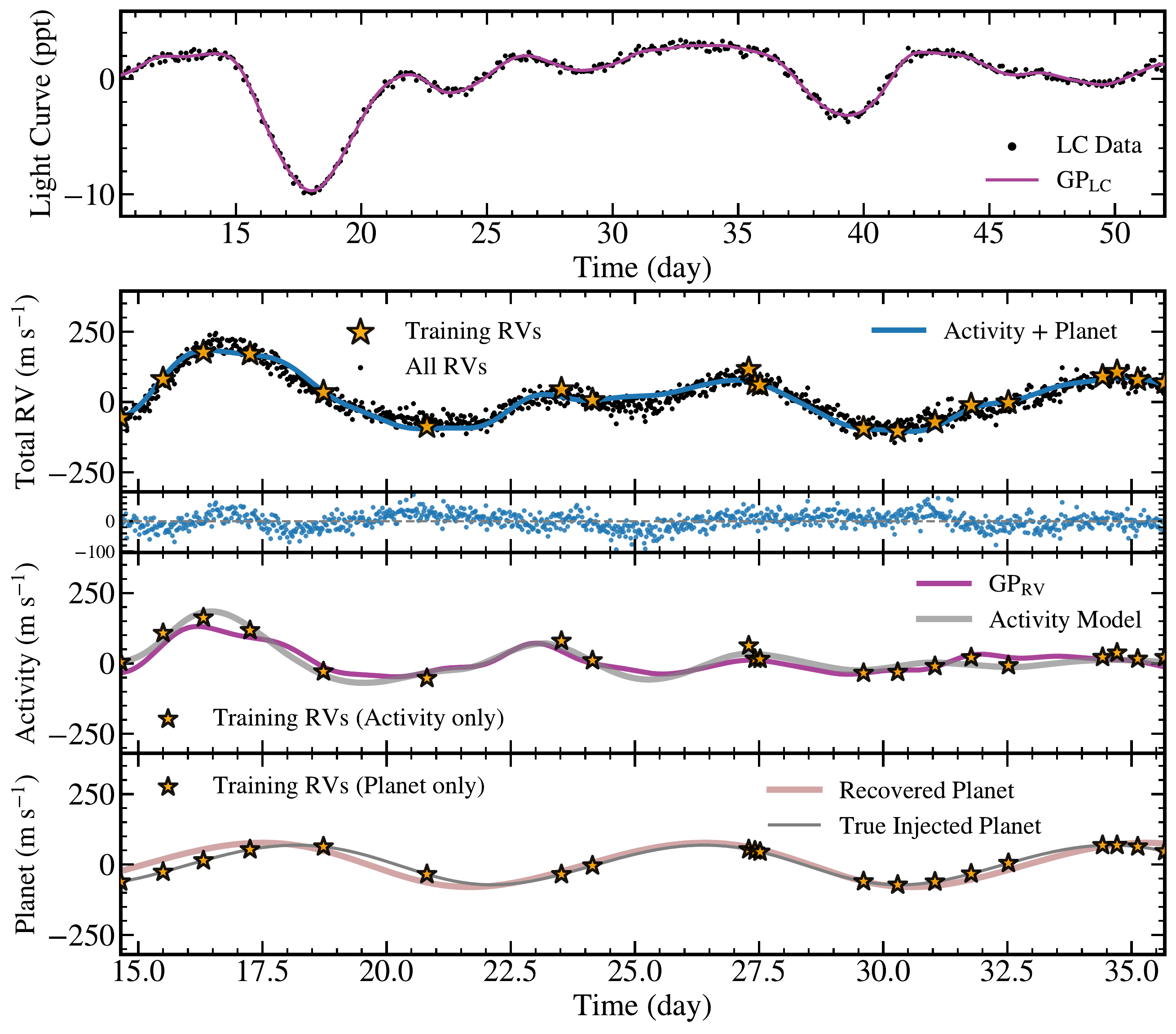}
    \caption{Mean prediction of the joint two latent GP stellar activity and Keplerian model. The photometric and RV times series displayed (black points and yellow stars in top two panels, respectively) are the same as in \autoref{fig:quasi_per_model_ex}. The joint model and accompanying residuals are shown in blue in the second and third panels. The individual stellar activity (purple) and the Keplerian (pink) models are plotted in the fourth and fifth panels, respectively. The training RVs, with only injected noise and individual activity and planetary components are also displayed as yellow stars. The model is able to predict the variations of the combined RV time series while simultaneously separating the two individual components.}
    \label{fig:planet_predict}
\end{figure*}

\autoref{fig:planet_predict} shows the mean prediction of the joint model. The photometry and RVs used to train the model are taken from the same time indices as in \autoref{sec:model_comparison} and \autoref{fig:model_predict}. The LC is displayed in the top panel and the total RV model, residuals, and individual components, stellar activity and Keplerian, are shown in the bottom panels. The model is able to predict the total variations in the RV dataset while simultaneously distinguishing between the contributions from the stellar activity and the planet. The model prediction performs the worst where the amplitude of the stellar activity ($\gtrsim$150 m s$^{-1}$) is over twice that of the injected planetary signal ($\sim$70 m s$^{-1}$). However, it is still able to place strong constraints on the total RV variations. For windows where the two components are equal in strength, the model fully captures both the stellar activity variations and the Keplerian curve. This result demonstrates the possibility of using the two latent GP model to distinguish between stellar activity and Keplerian signals in time series of systems resembling young, active stars harboring close-in giant planets. We note that future applications of this joint model should be cautious in instances where stellar activity overwhelmingly dominates the RV variations.

    \begin{deluxetable*}{lccc}[!t]
    \setlength{\tabcolsep}{10pt}
    \tablecaption{The injected Keplerian orbital parameters, the priors imposed during the sampling for each parameter, and the median values and $\pm 1\sigma$ HDI of each recovered posterior. \label{tab:planet_params}}
    \tablehead{\colhead{Parameter} & \colhead{Imposed Prior\tablenotemark{a}} & \colhead{True Injected Value} & \colhead{Recovered Value}}
    \startdata
    $P_\mathrm{orb} \: (d)$ & $\mathcal{U}[ 4, 50]$ & $8.2$ & $8.9^{+0.7}_{-0.9}$ \\
    $T_0 \: (d)$ & $\mathcal{U}[\mathrm{min}(t_\mathrm{RV}), \; \mathrm{max}(t_\mathrm{RV})]$ & $21.0$ & $20.1^{+2.7}_{-2.1}$ \\
    $K_p \: (m \: s^{-1})$ & $\mathcal{U}[50, 500]$ & $70.5$ & $78.6^{+10.0}_{-27.5}$ \\
    $\sqrt{e}\:\mathrm{sin}\:\omega$ & $\mathcal{U}[-1, 1]$ & $0.16$ & $0.21^{+0.37}_{-0.36}$ \\
    $\sqrt{e}\:\mathrm{cos}\:\omega$ & $\mathcal{U}[-1, 1]$ & $-0.12$ & $-0.08^{+0.38}_{-0.32}$ \\
    $e$ & $\mathcal{U}[0, 1)$ & $0.04$ & $0.22^{+0.12}_{-0.22}$ \\
    $\omega \: (\degree)$ & $\mathcal{U}[0, 2\pi]$ & $127$ & $134^{+76}_{-102}$ \\
    $i_* \: (\degree)$\tablenotemark{b} & $\cdots$ & $86.5 \degree$ & $\cdots$ \\ 
    \enddata
    \tablenotemark{a}{$\mathcal{U}[a, b]$ refers to the uniform distribution bounded by lower value $a$ and upper value $b$.}
    \tablenotetext{b}{Stellar inclination is previously defined as part of the stellar surface map initialization in \texttt{starry}. While the planet should transit due to the edge-on inclination and relatively low stellar obliquity ($\Psi_* = 6.0\degree$), we choose to not inject a transit model in order to simulate a non-transiting planet around a young, active star.}
    \end{deluxetable*}

\section{Discussion}
\label{sec:discuss}

\subsection{Limitations of GPs}

We note here that there are several caveats to using GPs, including several implications particular to our two latent GP framework. In general, GPs are not all powerful tools and careful consideration should be made on when to employ them. While our framework can be used for other time series data, such as spectroscopic activity indicators, the primary aim of our model is to improve on several weaknesses of GP implementation specifically for joint fitting of photometric and RV time series. Even in this use case, there still remains room for improvement and potential avenues for investigation.

\subsubsection{Variations in Observations}

The practical tests conducted in this study demonstrate that the two latent GP approaches can perform better at jointly modeling both simultaneous and non-simultaneous RV and photometric data as compared to other models. However, the extent to and parameter space over which this remains true is still unexplored. For specific use cases, these results are likely to vary. In \autoref{sec:model_performances}, we found that the one latent GP performed worse, likely due to the constraint from the greater number of photometric data points. Increasing the number of RVs or decreasing the number of photometric data points may change these results. Another adjustment is changing the cadence of the RV data. If there are large gaps in the RV observational coverage, whether the ancillary data is simultaneous, our GP model may not predict RV behavior as well as in the tests presented here. Similarly, as we examined in the gaps of the simultaneity test (\autoref{sec:non_simultaneous}), both the one and two latent GP frameworks continue to have strong predictive power even outside the photometric observational window. This predictive ability disperses some time after the contemporaneous coverage ends. However, how long this effect remains, how characteristics of stellar activity such as amplitude and spot decay time influence it, and how these parameters can affect model performance is still unknown.

A full investigation of these various changes to our approach and tests are needed. The computational cost required to fully explore this parameter space limits the number of tests conducted in this introduction of our framework. While we expect that over many simulations these experimental results will prove robust, a detailed study and further application of this model is critical and follows naturally from this work.

\subsubsection{Computational Cost and Scalability}

One well known drawback of employing GPs is the prohibitive computational cost, or dimensionality. GP regression requires matrix inversion, which scales as $\mathcal{O}(N^3)$, where $N$ is the number of data points. This concern is especially problematic for our framework, as we use both photometric and RV observations. While the number of observations is still tractable with only RVs, photometric data is often densely sampled and can reach tens of thousands of data points or more. One workaround, as done in this study, is to downsample or bin the LC, but this may not be feasible or desired.

Recently, much work has been done to scale Gaussian processes for large datasets and multiple dimensions by speeding up matrix inversion. \citet{Ambikasaran2015} demonstrated that for many standard covariance kernels, such as the squared exponential and Mat\'ern family of kernels (\autoref{sec:kernels}), a fast matrix inversion requiring only $\mathcal{O}(N\:\mathrm{log}^2N)$ operations is possible. This algorithm is implemented in the open software package \texttt{george}. For large 1D datasets, the \texttt{celerite} \citep{Foreman-Mackey2017, Foreman-Mackey2018} package applies GP regression with computational costs that scale as $\mathcal{O}(NJ^2)$, where $J$ is the number of terms in the covariance function. Inspired by \citet{Foreman-Mackey2017} and \citet{Gordon2020}, the open software package \texttt{S+LEAF 2} recently extended these scalable GP methods to multidimensional time series data \citep{Delisle2020, Delisle2022}. \texttt{S+LEAF 2} is able to jointly model multiple time series observations using GPs with computational costs that only scale linearly with $N$. With a sensible kernel choice, it is possible to scale up our GP framework to use the full LC and RV information with a reasonable computational cost employing these algorithms and packages.

\subsubsection{Framework Flexibility}

There are several physical justifications for increasing the number of latent processes in a joint GP model, some of which were presented in \autoref{sec:model}. For the purposes of simultaneously fitting photometric and RV time series, we considered here only two latent processes. In our case, we assumed that spot-like features on the stellar surface are the dominant source of activity and from a natural extension of the \textit{FF}$^\prime$ method arrived at utilizing two latent GPs. However, there are likely more complicated underlying physical processes that drive similarly complex relationships between RVs and other spectroscopic and photometric observations. For instance, these processes may include granulation and stellar pressure-mode, or “p-mode,” oscillations. These mechanisms evolve rapidly, on the timescales of minutes, and do not manifest apparent signals in ancillary times series such as photometry \citep[e.g.,][]{Luhn2023}. As such, these relationships likely cannot be modeled with a GP framework that uses only one process, two processes, or even combinations of those processes and their time derivatives. This is especially likely when we consider stellar hosts with different evolutionary stages and alternative dominant sources of activity.

GP frameworks jointly modeling more than two photometric and spectroscopic time series may likely benefit by increasing the number of latent processes. \citetalias{Rajpaul2015} and \citet{Barragan2022} note that different time series depend on the behavior of active regions in different ways. Some time series, like photometry or certain activity indicators such as log$\:R^\prime_\mathrm{HK}$ and S$_\mathrm{HK}$, depend only on the active region area (e.g., by only the latent process which describes this projected area). Other observables, such as RVs and the BIS, will be affected by both the active region and its time evolution (e.g., linear combinations of the GP and its time derivatives). The relationship between these traditional activity indicators and active regions has been previously investigated. However, spectral indices that have not been previously investigated may not fall within this dichotomy and require a more flexible framework in order to be jointly modeled.

Indeed, this was the case for \citet{Jones2017}, which generalized the work of \citetalias{Rajpaul2015} to include generic activity indicators, the second order time derivative, and an additional GP component to account for structured noise. Despite already accounting for second-order effects through higher time derivative terms, \citet{Jones2017} found that for different collections of spectroscopic time series, the additional process sometimes improved their model flexibility and performance. This could be further improved by continuing to increase the number of GPs while also relaxing the requirement of contemporaneous observations.

We expect that these arguments also apply to our two latent GP framework, but it becomes more difficult to justify increasing model complexity and potentially over-fitting the data. The addition of a second latent process is motivated as a means to fully capture the relationship between photometry and RVs described by the \textit{FF}$^\prime$ method. Despite the ideas listed above, there has not yet been a systematic study examining the possibilities of three or more latent GPs. A natural extension of this work would be to explore this.

\subsection{Targeting Young, Active Stars} \label{sec:young_stars}

The level of injected stellar activity in our simulations mimics the amplitudes observed in young, active stars. This choice is made in part to demonstrate the potential of the framework in the case of young, active stars. In particular, this level of RV jitter ($\sim$100 m s$^{-1}$) is expected for stars at an approximate age of 100 Myr ($10\mathrm{s}-100\mathrm{s}$ Myr) \citep[e.g.,][]{Tran2021}. The detection and characterization of planets in this age range, after the protoplanetary disk dissipates \citep[a few to 10 Myr;][]{Fedele2010, Barenfeld2016} but before most evolutionary process have completed ($< 1$ Gyr), is critical to our understanding of how planets evolve. Planets in this age range are important laboratories that can test our theories on processes that sculpt planet demographics such as orbital migration, atmospheric loss, and tidal realignment.

Several phenomena have characteristic timescales on the order of a couple 100 Myrs and therefore can be observationally distinguished by characterizing young planets. Giant planet orbital migration, for example, typically occurs on either $\lesssim 10$ Myr \citep[e.g.,][]{Ward1997, Kley2012, Albrecht2012} or $10^7$--$10^{10}$ yr timescales, depending on whether in-spiraling disk migration or three-body dynamical interactions with an outer companion plays the dominant role. Similarly, recent tidal erasure simulations predict that spin-orbit realignment of a hot Jupiter ($P_\mathrm{orb} \lesssim 10$ d) orbiting a cool star \citep[below the Kraft break, $T_\mathrm{eff} \gtrsim 61250$ K][]{Kraft1967} is not possible if the planet migrated inward after 100 Myr. Searching for young, giant planets can allow us to infer the dominant migration mechanism sculpting giant planet architectures.

Atmospheric mass loss is another process we can investigate by studying young planets. The observed paucity of intermediate-sized planets called the radius valley \citep[$1.5 - 2 \; R_\oplus$;][]{Fulton2017} has been connected to atmospheric loss mechanisms. The two most prevalent mass loss models are photoevaporation and core-powered mass loss. The former predicts most mass loss to occur within the first few hundred Myrs \citep[e.g.,][]{Owen2017}, while the latter suggests a timescale of a few Gyr \citep[e.g.,][]{Gupta2019}. Searching for young super-Earths around this age could potentially reveal the dominant thermal evolutionary pathway these stripped core planets undertake \citep[e.g.,][]{Lopez2013, Jin2018, Berger2020, Sandoval2021, David2021, Gupta2021, Rogers2021}.

Observations of young planets are already challenging our current theories of planet formation. A notable example is the V1298 Tau system \citep[$\sim 20$ Myr;][]{David2019}. Core accretion models for giant planet formation predict that gaseous planets will contract as they cool \citep[e.g.,][]{Fortney2007, Mordasini2012}. However, \citet{Suarez2022} infer masses for V1298 Tau b and e that are significantly larger than expected for this age. This may suggest that the V1298 Tau system underwent much more rapid contraction than expected. This picture of thermal evolution is further complicated by the observed abundance of young planets with radii larger than planets with similar characteristics but at older ages \citep[e.g.,][]{Obermeier2016, Pepper2017, Mann2018, Newton2019}. Robustly measuring the masses and radii of young planets may help reveal what drives planetary mass loss and contraction.

Based on the tests presented here, the two latent GP framework seems particularly effective at mitigating spot-driven activity signals--signals that dominate in the young star regime. Further investigation on the applicability of this framework to real time series data of young, active stars may prove especially fruitful but are still needed.

\section{Summary}
\label{sec:summary}

We have presented a new GP framework designed to jointly model activity signals in RV and photometric time series data. Our approach is inspired by the \textit{FF}$^\prime$ method introduced by \citet{Aigrain2012}. In order for a GP framework to jointly model RVs and photometry in a similar fashion as the \textit{FF}$^\prime$ model, a minimum of two or more latent processes is required. Our approach is the first to jointly model both time series as the combination of two latent processes and their time derivatives in a way that can account for every term in the \textit{FF}$^\prime$ relationships.

We compare our GP framework with approaches adopted in the literature using synthetic photometry and RV time series generated with the \texttt{starry} \citep{Luger2019} open software package. We introduce an evolving starspot model that we implement in \texttt{starry} to generate light and RV curves with realistic observational cadence. We apply our GP framework and other approaches previously used in the literature to synthetic datasets generated using this new starspot model. For a sensible comparison between representative model fits, we find that our GP model outperforms other approaches in predicting RV variations caused by activity features. We also find that LCs do provide information on RV variations, even when the two time series are non-contemporaneous. This asynchronous constraining power is stronger when the photometry is observed prior to the RV measurements. Even with varying simultaneity, we see that our model still predicts spot-driven RV behavior better than the other frameworks.

In the future, this work can be further developed by using our GP framework to conduct a systematic study on the parameter space over which jointly modeling activity signals in photometry and RVs is most effective. Potential extensions could include investigating the limits of non-simultaneity between photometric and RV time series, the optimum RV observational cadence needed to constrain stellar activity, and the applicability of this joint framework on facilitating the detection and mass determinations of exoplanets.

Finally, we note that the last of these potential applications, specifically the detection of young planets, can be especially fruitful. We demonstrate, in a limited capacity, that the two latent GP model can be used to jointly fit stellar activity and a Keplerian signal. The model is able to predict the total RV variations as well as distinguish between the two components. The tests applied here exhibit the versatility of our framework for stars exhibiting a high level of spot-driven activity. Mitigation of these signals through modeling can increase RV sensitivity and ultimately facilitate the characterization of young planets. These systems can then be leveraged to test our current theories on evolution processes such as orbital migration and atmospheric loss.

\begin{acknowledgements}

\software{\texttt{emcee} \citep{Foreman-Mackey2013, Foreman-Mackey2019b}, \texttt{starry} \citep{Luger2019}, \texttt{celerite2} \citep{Foreman-Mackey2017, Foreman-Mackey2018}, \texttt{astropy} \citep{Astropy2018}, \texttt{matplotlib} \citep{Hunter4160265}.}

\end{acknowledgements}

We thank Trevor David, David Hogg, Oscar Barrag{\'a}n, Christian Gilbertson, Belinda Nicholson, and Suzanne Aigrain for their helpful insights on all things GP, stellar activity, photometry, and RVs. We also thank the referee for their helpful comments.

Q.H.T. acknowledges the support from a NASA FINESST grant (80NSSC20K1554).

\appendix

\section{Covariance of a Latent Gaussian Process and its Time Derivative}
\label{sec:covariance_time_derivative}

We want to model time series observations as linear combinations of some latent GP (or set of GPs) and its time derivatives. This model is defined by the covariance block matrix, which is constructed from the covariance functions between the various time series observation. These covariance functions can be related to the covariance between an observation of a GP and its time derivative, which we consider here.

If a process is defined as $f(t)$ and its time derivative as $\dot{f}(t)$, then the covariance between an observation of the process at time $t_\mathrm{n}$ with itself at time $t_\mathrm{m}$ can be represented by: 
\begin{equation} \label{eq:gp_cov1}
    \mathrm{cov}\Big[f(t_\mathrm{n}), f(t_\mathrm{m})\Big] = k(t_\mathrm{n}, t_\mathrm{m}).
\end{equation}
The covariances between observations of the process $f$ and $\dot{f}$ at times $t_\mathrm{n}$ and $t_\mathrm{m}$ are thus:
\begin{equation} \label{eq:gp_cov23}
    \begin{split}
    \mathrm{cov}\Big[f(t_\mathrm{n}),\dot{f}(t_\mathrm{m})\Big] &= \frac{dk(t_\mathrm{n}, t_\mathrm{m})}{dt_\mathrm{m}}, \\
    \mathrm{cov}\Big[\dot{f}(t_\mathrm{n}), f(t_\mathrm{m})\Big] &= \frac{d k(t_\mathrm{n}, t_\mathrm{m})}{d t_\mathrm{n}},
    \end{split}
\end{equation}
which come directly from taking the derivative of \autoref{eq:gp_cov1} with respect to the corresponding time index:
\begin{equation}
    \begin{split}
        \frac{d}{dt_\mathrm{m}}\Big[ k(t_\mathrm{n}, t_\mathrm{m}) \Big] &= \frac{d}{dt_\mathrm{m}}\bigg\{\mathrm{cov}\Big[f(t_\mathrm{n}), f(t_\mathrm{m})\Big]\bigg\}, \\
        &= \frac{d}{dt_\mathrm{m}} \bigg\{\mathrm{E}\Big[ f(t_\mathrm{n}) \cdot f(t_\mathrm{m}) \Big]\bigg\}, \\
        &= \mathrm{E} \bigg[ f(t_\mathrm{n}) \cdot \frac{  df(t_\mathrm{m})}{dt_\mathrm{m}} + f(t_\mathrm{m}) \cdot \frac{df(t_\mathrm{n})}{dt_\mathrm{m}}  \bigg], \\
        &= \mathrm{E} \bigg[ f(t_\mathrm{n}) \cdot \frac{  df(t_\mathrm{m})}{dt_\mathrm{m}}  \bigg], \\
        &= \mathrm{E} \Big[ f(t_\mathrm{n}) \cdot \dot{f}(t_\mathrm{m}) \Big], \\
        &= \mathrm{cov} \Big[ f(t_\mathrm{n}), \dot{f}(t_\mathrm{m}) \Big].
    \end{split}
\end{equation}
Similarly, the covariance between observations of the time derivative with itself at times $t_\mathrm{n}$ and $t_\mathrm{m}$ is defined as:
\begin{equation} \label{eq:gp_cov4}    
    \mathrm{cov}\Big[\dot{f}(t_\mathrm{n}), \dot{f}(t_\mathrm{m})\Big] = \frac{d^2 k(t_\mathrm{n}, t_\mathrm{m})}{d t_\mathrm{n} d t_\mathrm{m}}.
\end{equation}

\section{Elements of the Block Matrix}
\label{sec:block_matrix_elements}

We can define the covariance function of different combinations of F and $\Delta$RV in terms of the covariance for each of the terms in their expression as in \autoref{eq:K11}. Using the definitions of F and $\Delta$RV in \autoref{eq:flux} and \autoref{eq:delta_rv} and the notation described in \autoref{sec:covariance_time_derivative}, the elements in the block matrix of our GP framework (\autoref{eq:block_matrix}) are

\begin{subequations}
\begin{align}
    \begin{split} \label{eq:K12}
        K_\mathrm{F,RV}(t_\mathrm{F}, t_\mathrm{RV}) &= AE\, k_f(t_\mathrm{F}, t_\mathrm{RV}) + AH\, \frac{dk_f(t_\mathrm{F}, t_\mathrm{RV})}{dt_\mathrm{RV}} + BE\, \frac{dk_f(t_\mathrm{F}, t_\mathrm{RV})}{dt_\mathrm{F}} + BH\, \frac{d^2k_f(t_\mathrm{F}, t_\mathrm{RV})}{dt_\mathrm{F}dt_\mathrm{RV}}  \\
        &\qquad + CI\, k_g(t_\mathrm{F}, t_\mathrm{RV}) + CJ\, \frac{dk_g(t_\mathrm{F}, t_\mathrm{RV})}{dt_\mathrm{RV}} + DI\, \frac{dk_g(t_\mathrm{F}, t_\mathrm{RV})}{dt_\mathrm{F}} + DJ\, \frac{d^2k_g(t_\mathrm{F}, t_\mathrm{RV})}{dt_\mathrm{F}dt_\mathrm{RV}},
    \end{split}
    \\[2ex]
    \begin{split} \label{eq:K21}
        K_\mathrm{RV, F}(t_\mathrm{RV}, t_\mathrm{F})
        &= EA\, k_f(t_\mathrm{RV}, t_\mathrm{F}) + EB\, \frac{dk_f(t_\mathrm{RV}, t_\mathrm{F})}{dt_\mathrm{F}} + HA\, \frac{dk_f(t_\mathrm{RV}, t_\mathrm{F})}{dt_\mathrm{RV}} + HB\, \frac{d^2k_f(t_\mathrm{RV}, t_\mathrm{F})}{dt_\mathrm{RV}dt_\mathrm{F}} \\
        &\qquad + IC\, k_g(t_\mathrm{RV}, t_\mathrm{F}) + ID\, \frac{dk_g(t_\mathrm{RV}, t_\mathrm{F})}{dt_\mathrm{F}} + JC\, \frac{dk_g(t_\mathrm{RV}, t_\mathrm{F})}{dt_\mathrm{RV}} + JD\, \frac{d^2k_g(t_\mathrm{RV}, t_\mathrm{F})}{dt_\mathrm{RV}dt_\mathrm{F}},
    \end{split}
    \\[2ex]
    \begin{split} \label{eq:K22}
        K_\mathrm{RV,RV}(t_\mathrm{RV}, t_\mathrm{RV}') &= E^2\, k_f(t_\mathrm{RV}, t_\mathrm{RV}') + EH\, \frac{dk_f(t_\mathrm{RV}, t_\mathrm{RV}')}{dt_\mathrm{RV}'} + HE\, \frac{dk_f(t_\mathrm{RV}, t_\mathrm{RV}')}{dt_\mathrm{RV}} + H^2\, \frac{d^2k_f(t_\mathrm{RV}, t_\mathrm{RV}')}{dt_\mathrm{RV}dt_\mathrm{RV}'} \\
        &\qquad + I^2\, k_g(t_\mathrm{RV}, t_\mathrm{RV}') + IJ\, \frac{dk_g(t_\mathrm{RV}, t_\mathrm{RV}')}{dt_\mathrm{RV}'} + JI\, \frac{dk_g(t_\mathrm{RV}, t_\mathrm{RV}')}{dt_\mathrm{RV}} + J^2\, \frac{d^2k_g(t_\mathrm{RV}, t_\mathrm{RV}')}{dt_\mathrm{RV}dt_\mathrm{RV}'}.
    \end{split}
\end{align}
\end{subequations}

\section{Conditional Predictions of Other GP Kernels} \label{sec:app_other_kernels}

\begin{figure*}[!th]
    \centering
    \includegraphics[width=1\linewidth]{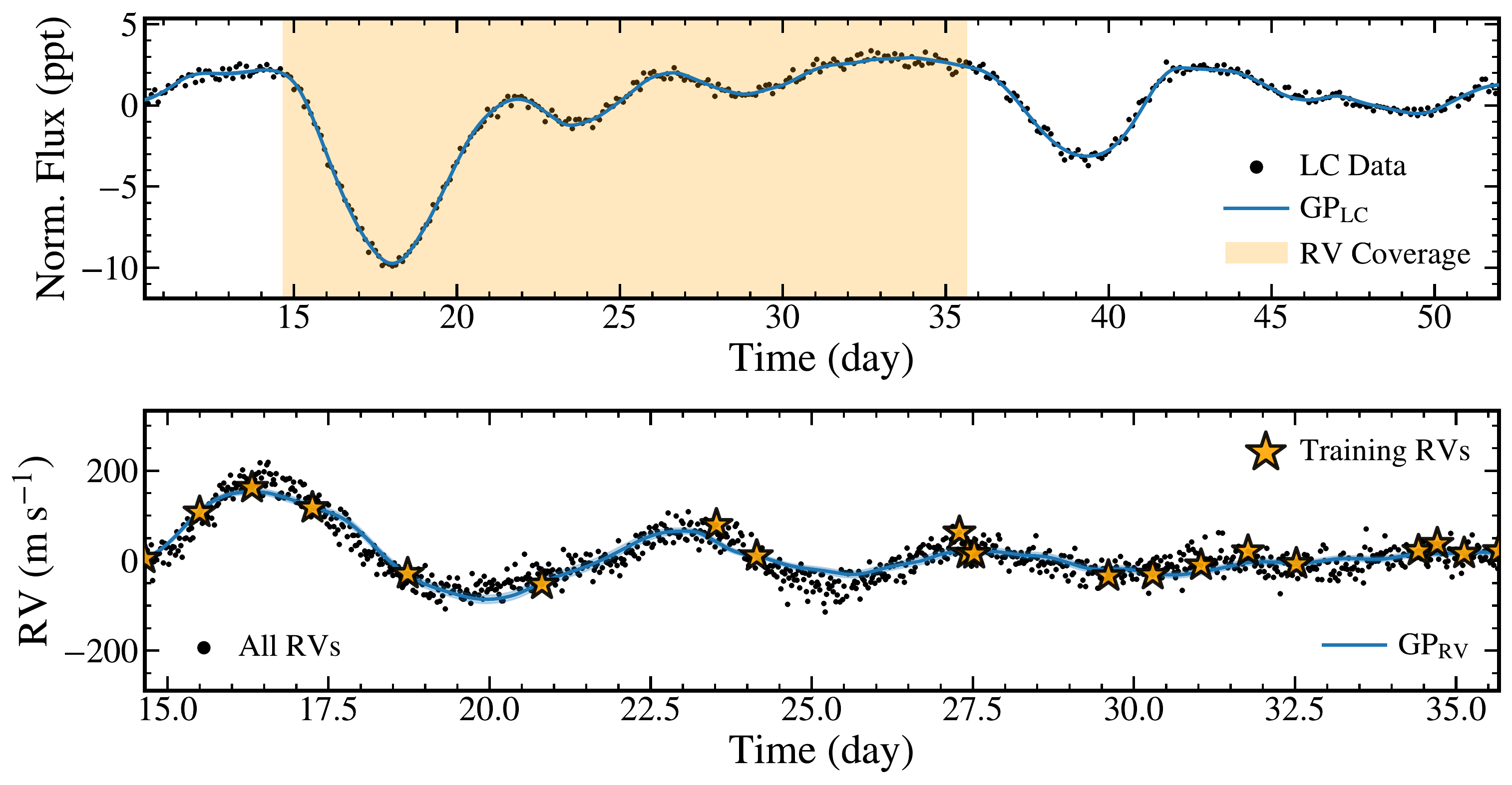}
    \caption{The same application of our two independent latent GP framework as in \autoref{fig:quasi_per_model_ex} on the fiducial synthetic dataset using a Mat\'ern-5/2 kernel function.}
    \label{fig:matern52_model_ex}
\end{figure*}

\begin{figure*}[!th]
    \centering
    \includegraphics[width=1\linewidth]{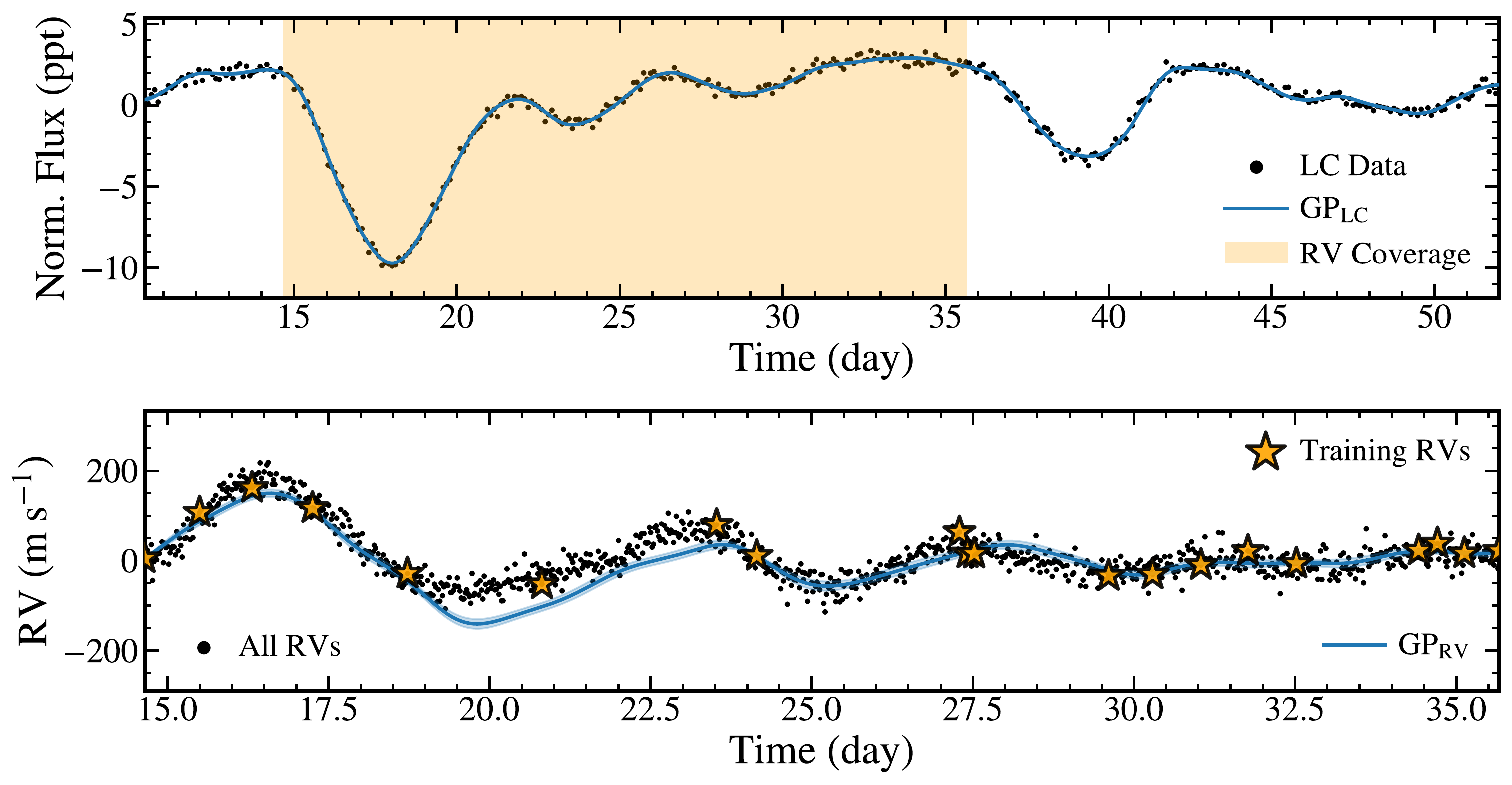}
    \caption{The same application of our two independent latent GP framework as in \autoref{fig:quasi_per_model_ex} on the fiducial synthetic dataset using an SE kernel function.}
    \label{fig:sq_exp_model_ex}
\end{figure*}

Any covariance kernel function that meets the criteria specified in \autoref{sec:kernels} can be used in the two latent GP framework. Two kernels that are explicitly defined are the M$\sfrac{5}{2}$ and SE kernels. \autoref{fig:matern52_model_ex} and \autoref{fig:sq_exp_model_ex} show the GP regression with the two latent framework using these two kernel functions, respectively, for the fiducial time series.

\bibliography{gp}{}
\bibliographystyle{aasjournal}

\end{document}